# Molecular notation can act as a shortcut in AI benchmark evaluation

Yosuke Kikuchi[1], Yasuhiro Yoshikai[1], Shumpei Nemoto[1], Ayako Furuhama[2], Takashi Yamada[3], Hiroyuki Kusuhara[1], Tadahaya Mizuno[1, 4, †]

[1] Laboratory of Molecular Pharmacokinetics, Graduate School of Pharmaceutical Sciences, The University of Tokyo, 7-3-1 Hongo, Bunkyo, Tokyo, Japan

[2] Division of Genome Safety Science, National Institute of Health Sciences, 3-25-26 Tonomachi, Kawasaki-ku, Kawasaki, Kanagawa 210-9501, Japan

[3] Division of Cellular and Molecular Toxicology, National Institute of Health Sciences, 3-25-26 Tonomachi, Kawasaki-ku, Kawasaki, Kanagawa 210-9501, Japan

[4] The Institute of Statistical Mathematics (ISM), Research Organization of Information and Systems, Tachikawa, Tokyo 190-8562, Japan

† Correspondence: Tadahaya Mizuno (tadahaya@gmail.com)

## Abstract

High predictive accuracy does not necessarily mean that an AI model relies on information relevant to its scientific task. Molecular prediction illustrates this concern: the same molecule can be represented by different SMILES strings, potentially altering model behavior. We combined a systematic PubMed literature survey with experiments on chemical language models. SMILES preparation could not be identified from articles or Supplementary Information in 71 of 174 studies (40.8%). Across six benchmarks, 58.7% of identity-preserving input pairs used different SMILES strings. Notation differences changed reconstruction and latent representations, while separately trained classifiers often achieved similar scores. In two datasets, notation was associated with labels and changed predictions of a fixed model, showing how benchmark scores can reward notational shortcuts. Balancing notation during classifier training reduced prediction differences but sometimes lowered performance on original inputs. These findings support evaluating predictive performance alongside consistency under identity-preserving changes in input representation.

**Short title**: Molecular notation and AI benchmarks

When a model predicts a molecular property, the intended object is the molecule, not the notation used to describe it. Under the same conditions, alternative descriptions that preserve molecular identity should therefore support consistent predictions. Chemical language models (CLMs) use molecular strings like SMILES (Simplified Molecular Input Line Entry System) as a molecular input representation. The same molecule can have several valid SMILES strings, which may differ in atom order, branching, ring traversal, or aromatic bond notation[1,2]. Canonicalization selects one string according to a particular algorithm, but different toolkits and settings can produce different canonical strings for the same molecule [3–5]. These alternatives describe the same molecular structure but give a sequence model different inputs.

Previous studies have addressed several aspects of this problem. Chemical structure preparation can affect cheminformatics analyses[6]. Randomized-SMILES augmentation[7] and molecular translation[8] use alternative strings to improve learning. Prior work has also assessed robustness to identity-preserving SMILES transformations, including Kekulization, in both molecular retrieval and downstream prediction[9]. More broadly, shortcut learning and underspecification show why strong test performance need not identify the decision rules researchers expect a model to use[10,11]. The question is therefore not only whether models respond to notation, but whether naturally occurring notation patterns can influence benchmark scores and favor notation-dependent predictions.

This raises a question of alignment between the intended scientific task and model behavior. Here, alignment means respecting the distinctions and equivalences that define the task, rather than reproducing human reasoning. For molecular property prediction, consistency under identity-preserving notation changes is one such requirement. This consistency may be achieved by a documented preprocessing procedure and need not arise within the encoder alone. Reproducing the input procedure and testing the resulting predictions are therefore complementary requirements. A high benchmark score alone establishes neither.

We first surveyed 174 studies to assess how SMILES preparation is documented in practice. For 61 studies (35.1%), no description of SMILES preparation was found in the reviewed sources, including linked repositories where available. We then investigated how notation differences affect molecular reconstruction, latent representations, and property prediction in public benchmarks[12,13]. We further examined whether notation patterns were associated with labels, whether changing notation altered predictions of a fixed model, and whether balanced classifier training reduced these effects. Additional model architectures were used to examine whether selected findings extended beyond the primary model.

The results show three ways in which notation can affect evaluation. Its effects can be visible in reconstruction and latent representations, masked when classifiers are fitted separately to each SMILES input condition, or rewarded when a label-associated notation pattern improves prediction. We refer to these as visible, masked, and rewarded sensitivity. Their significance extends beyond sensitivity to SMILES: benchmark evaluation can favor behavior that conflicts with the equivalences the scientific task requires. Molecular identity thus provides a concrete basis for testing this aspect of alignment independently of predictive performance.

## Results

### Reporting of model-ready SMILES preparation

The PubMed search on 11 August 2026 yielded 353 records. After excluding 78 records during title and abstract screening, we assessed 275 full texts and excluded a further 101, leaving 174 included studies (Supplementary Methods S1 and Supplementary Data 1).

SMILES preparation was classified from the article or Supplementary Information for 103 of the 174 studies (59.2%). For the remaining 71, we checked whether a publication-linked GitHub repository was available and examined its contents when accessible. Additional preparation information was found for 10 studies (5.7% of all included studies); no relevant description was identified for 61 (35.1%) (Fig. 1B).

Thus, SMILES preparation could not be identified from the article or Supplementary Information in 71 studies (40.8%) and remained unresolved in 61 (35.1%) after repository follow-up. Importantly, this finding concerns reporting; it does not imply that preprocessing was absent or that an identified description provides every setting needed for reproduction.

### Public benchmarks contain different SMILES strings for the same molecules

The MoleculeNet source tables[12] contained 22,901 records across BACE, BBBP, ClinTox, SIDER, Tox21, and ToxCast. The retained model inputs contained 22,784 records; 117 source records were absent, and the inclusion mask was fixed independently of task-label availability (Supplementary Table S1; Supplementary Data 2). Of the dataset-provided/RDKit-harmonized pairs, 22,776 preserved molecular identity. Among these, 13,366 (58.7%) used different strings (Fig. 1C). The fraction ranged from 2.8% in Tox21 to 100% in BACE, showing that notation variation differed greatly among datasets.

Different benchmark-provided strings can describe the same molecule: ketamine provides one example across four MoleculeNet datasets (Fig. 1D). Representative dataset-provided and RDKit-harmonized pairs further illustrate changes in branch placement, ring traversal, and Kekulé/aromatic notation (Supplementary Fig. S1). Stereochemical completeness was examined separately: preserving the molecular identity encoded in a string does not require every potential stereochemical site to be specified (Supplementary Methods S2 and Supplementary Table S2). We therefore distinguished differences in notation from incomplete molecular specification and changes in chemical state[6].

Notation differences were common but unevenly distributed across datasets. We next tested whether their effects on model behavior were also dataset dependent. An analysis of 22 Therapeutics Data Commons (TDC) ADME datasets extended this comparison beyond MoleculeNet[13]. Among 90,866 parseable records after component selection, 2,806 (3.1%) changed after RDKit harmonization. Dataset-level fractions ranged from 0.6% to 66.2% (Supplementary Table S3). Thus, the low overall percentage concealed large differences among datasets.

## Notation changes molecular reconstruction and latent representations

To examine how notation affects molecular reconstruction, we used a model trained to translate randomized SMILES into canonical SMILES, following the representation-learning approach used for continuous and data-driven descriptors (CDDD)[8]. Our implementation used the Transformer-based variational autoencoder (Transformer–VAE) architecture described previously[14] and was pretrained with RDKit-generated inputs and targets. Both benchmark input conditions were evaluated with the same fixed model and decoding procedure. RDKit-harmonized strings therefore matched the training convention more closely.

The size of the difference was substantial in five datasets: harmonization increased connectivity-identity reconstruction by 0.282–0.754, with the largest increase in SIDER, from 0.192 to 0.947 (Fig. 2A). All five paired 95% confidence intervals excluded zero. Tox21 showed little difference (0.981 versus 0.981), consistent with its low changed-string fraction. Exact-sequence and stereo-aware identity reconstruction followed the same dataset-level pattern (Supplementary Results S1 and Supplementary Fig. S2).

Notation also changed the latent representations. Dataset-provided and harmonized inputs occupied partly distinct regions in a joint t-SNE projection of SIDER (Fig. 2B), and distances between paired vectors in the full latent space varied across datasets (Fig. 2C). Among pairs with different strings but the same molecular identity, larger string differences were associated with larger latent differences in every dataset (Spearman $\rho$=0.10–0.74; Supplementary Results S2 and Supplementary Fig. S3). The same molecule therefore did not always receive the same internal representation.

A variational autoencoder based on gated recurrent units (GRU–VAE) showed the same reconstruction pattern in the three datasets that passed the unchanged-input consistency check: harmonization improved reconstruction in BBBP and ClinTox, while Tox21 remained effectively unchanged (Supplementary Fig. S4 and Supplementary Table S4). This result extends the observed reconstruction sensitivity to a recurrent model within the tested subset.

## Similar predictive performance can hide notation sensitivity

Predicting properties such as biological activity and toxicity is a major downstream application of CLMs[8,15]. For property prediction, the fixed Transformer–VAE encoder generated latent vectors for both input conditions. We then trained a separate XGBoost classifier[16] for each condition using the same train/test split (Fig. 3A). This comparison asks whether each representation can support useful predictions after classifier training. It does not test whether a single fixed predictor gives the same result for different notations.

Test-set performance was similar in BACE, SIDER, Tox21, and ToxCast, with absolute point differences of at most 0.014 for AUROC and 0.024 for AUPR (Fig. 3B; Supplementary Table S5). Thus, large changes in reconstruction and latent representations could coexist with similar predictive performance when classifiers were trained separately.

Unexpectedly, ClinTox and BBBP showed the opposite pattern: although harmonization improved molecular reconstruction, property-prediction performance was higher with the dataset-provided SMILES. In BBBP, dataset-provided versus harmonized AUROC was 0.960 versus 0.897 (difference, +0.063; 95% CI, +0.034 to +0.094), and AUPR was 0.986 versus 0.957 (+0.029; +0.014 to +0.047). In ClinTox, dataset-provided versus harmonized AUROC was 0.967 versus 0.794 (difference, +0.173; 95% CI, +0.102 to +0.252), and AUPR was 0.891 versus 0.247 (+0.644; +0.470 to +0.767). This discrepancy prompted us to examine the information available in the two input forms. Since the molecular identities were unchanged, we considered whether notation conventions themselves might provide an additional predictive cue. We therefore compared notation patterns between positive and negative compounds.

**Kekulé/aromatic notation is associated with labels and affects fixed-model predictions**

We focused on a clear difference in aromatic-ring notation in ClinTox and BBBP. Some dataset-provided strings used Kekulé notation, with alternating single and double bonds, whereas the corresponding RDKit-harmonized strings used lowercase aromatic notation. We classified records showing this difference across all aromatic bonds as K/A records, provided that connectivity and encoded stereochemistry were preserved. The classification was defined without labels. Partial or mixed aromaticity patterns were also present, but were examined separately rather than included in this category (Supplementary Table S6).

The label association was strong. All 96 K/A records in ClinTox were positive, accounting for 85.7% of positive records and none of the negative records (two-sided Fisher exact $P=2.2\times10^{-134}$; Fig. 4A). In BBBP, 835 of 843 K/A records were positive (99.1%), covering 53.6% of positives but 1.7% of negatives ($P=7.0\times10^{-116}$; Fig. 4B).

We next tested whether K/A notation alone could distinguish positive from negative compounds in the test set. Each record received a score of 1 if it belonged to the K/A category and 0 otherwise; no molecular graph or learned embedding was used (Fig. 4C). In ClinTox, this binary indicator achieved AUROC 0.886 (95% CI, 0.795–0.977) and AUPR 0.790 (0.622–0.958), compared with a positive prevalence of 0.075. In BBBP, AUROC was 0.778 (0.749–0.808) and AUPR was 0.895

(0.881–0.909), compared with a positive prevalence of 0.765. These results concern the defined K/A pattern. Additional notation features and sensitivity analyses excluding molecular identities present in training are provided in the Supplementary Information.

To test whether the fixed model responded to this notation pattern, we used RDKit to generate Kekulé and aromatic strings from the same molecular structure. For each held-out K/A molecule, we compared predictions for the original benchmark string and these two RDKit-generated alternatives. The RDKit-generated Kekulé string served as a rewriting control: it retained Kekulé notation and gave probabilities similar to those from the original string. In contrast, changing the RDKit-generated Kekulé input to its aromatic counterpart reduced the positive-class probability in all 17 ClinTox pairs and in 176 of 178 BBBP pairs. The mean decrease was 0.455 (95% CI, 0.324–0.583) in ClinTox and 0.144 (0.127–0.162) in BBBP; 10 and 9 records, respectively, crossed the fixed decision threshold (Fig. 4D).

We next assessed how these prediction shifts affected benchmark performance by replacing only the held-out K/A strings with the RDKit-generated aromatic forms and leaving all other inputs unchanged. ClinTox AUROC decreased from 0.967 to 0.663 (Δ=−0.304; 95% CI, −0.421 to −0.193), and AUPR decreased from 0.891 to 0.185 (Δ=−0.706; −0.834 to −0.528). BBBP AUROC decreased from 0.960 to 0.873 (Δ=−0.087; −0.118 to −0.060), and AUPR decreased from 0.986 to 0.942 (Δ=−0.044; −0.065 to −0.026) (Fig. 4E). Together, label association, notation-only prediction, and this fixed-model comparison supported a contribution from label-associated notation.

Because 32 of the 195 initial RDKit-generated held-out pairs also differed in parsed atom order, we performed a second intervention in which Kekulé and aromatic strings were matched for atom and bond output order while preserving molecular identity. The fixed model assigned higher positive-class probabilities to K than A in 307 of 313 eligible BBBP records and 204 of 205 ClinTox records. Mean K-minus-A shifts were 0.161 (95% CI, 0.146–0.177) and 0.331 (0.298–0.365), respectively. After excluding identities present in training, the shifts remained 0.167 and 0.320. Prediction sensitivity therefore persisted after output-order matching (Supplementary Methods S6; Supplementary Table S7). This order-matched experiment tested notation dependence in a separate eligible population, whereas the initial selective intervention measured the change in benchmark performance when naturally occurring K/A records were rewritten.

**Notation dependence extends across pretrained encoders and is reduced by balanced training**

We next examined whether notation dependence also arose with a distinct pretrained encoder. We therefore compared the Transformer–VAE with MoLFormer, a large-scale pretrained CLM[15]. The shared dataset contained 2,559 records after tokenizer filtering and removal of identities with conflicting labels, corresponding to 1,470 BBBP and 1,033 ClinTox molecular groups. The Transformer–VAE and MoLFormer encoders were evaluated using identical group-separated folds across five seeds (Fig. 5; Supplementary Methods S8). Both encoders remained fixed. Balanced classifier training gave equal weight to K and A representations while preserving the total training weight per molecular identity.

We first evaluated classifiers trained on dataset-provided inputs using the original, strict Kekulé (K), and strict aromatic (A) input representations without refitting the classifiers (Fig. 5A,E). For MoLFormer, AUROC in BBBP was 0.953, 0.861, and 0.889 for original, strict K, and strict A inputs, respectively; the corresponding ClinTox values were 0.999, 0.694, and 0.601. The Transformer–VAE also showed substantial input-dependent changes. Strict K and A were matched for atom and bond output order, whereas comparisons involving the original strings may include additional serialization differences.

We next examined whether balanced K/A training reduced this dependence. For MoLFormer with XGBoost, balanced training reduced the mean absolute K/A probability difference from 0.217 to 0.057 in BBBP and from 0.570 to 0.026 in ClinTox. AUROC of K/A-averaged predictions increased from 0.891 to 0.934 and from 0.707 to 0.896, respectively, while AUROC on dataset-provided inputs decreased from 0.953 to 0.929 and from 0.999 to 0.811 (Fig. 5B–D,F–H). The Transformer–VAE showed the same pattern, and these directions held across all five seeds for both datasets. Logistic classifiers provided a complementary comparison (Supplementary Table S8).

Balanced classifier training reduced the difference between K and A predictions, but performance depended on which input forms were evaluated. K/A-averaged performance also includes the effect of averaging two predictions; single-K and single-A results and the random-assignment control are reported separately. High performance on dataset-provided inputs therefore did not imply low notation dependence, even with a different pretrained encoder. Per-seed metrics for all classifier-training and evaluation conditions, including strict K/A comparisons, are provided in Supplementary Data 3.

## Discussion

These findings identify a potential mismatch between the molecular properties that models are intended to predict and the behavior rewarded by benchmark evaluation. In two datasets, SMILES notation was associated with labels, and rewriting identity-preserving inputs changed fixed-model predictions; reducing this dependence could lower performance on the original inputs. In a systematic survey of 174 PubMed-indexed studies, SMILES preparation could not be identified from the article or Supplementary Information in 71 (40.8%). This reporting gap does not establish whether the surveyed models were notation sensitive, but it shows that input preparation is not consistently documented in the assessed literature. Together, the experimental and survey findings connect a measurable notational shortcut with a reporting problem relevant to the evaluation of molecular models.

Molecular identity provides a reference independent of both the model and its performance metric. By preserving connectivity and encoded stereochemistry, the controlled comparisons changed SMILES notation without changing the molecular identity represented by the input. This makes it possible to test whether a prediction depends on an aspect of the input representation that is irrelevant to the intended task. Different latent vectors alone do not establish a failure of alignment: a latent representation may retain notation information and still support consistent, useful predictions. More direct evidence comes from changes in the output of a fixed predictor for the same molecule under the same conditions. This criterion also distinguishes notation changes from changes in stereochemistry or chemical state, for which different predictions may be appropriate. Distinguishing different stereoisomers is a separate requirement, and difficulties in learning chirality have been reported for SMILES-based Transformers[17].

Shortcut learning and underspecification are established concerns in predictive modeling[10,11]. Our results reveal a specific form of this problem in molecular prediction: SMILES notation can carry label information while preserving molecular identity. AMORE assessed robustness to identity-preserving SMILES transformations, including Kekulization, in molecular retrieval and downstream prediction[9]. Here, a notation-only score distinguished positive from negative compounds, and identity-preserving changes altered predictions from a fixed model. The systematic literature survey provides complementary evidence that the preparation of model inputs is not always documented in the reviewed studies. This does not mean that the surveyed models relied on the shortcut, but it makes the possibility harder to evaluate from published benchmark results alone. The performance decrease after rewriting inputs remains specific to the tested intervention, not an exact decomposition of the score into chemical and notational contributions.

The balanced-training comparison has a practical consequence for model selection. When the encoders were kept fixed, giving both notations equal weight during classifier training reduced the difference between their predictions. Yet evaluation on the original benchmark inputs favored the more notation-dependent classifiers in the main comparisons. Improvement in a scientifically desirable property can therefore appear as deterioration under the original evaluation criterion. The appropriate response is not to replace predictive performance with invariance, but to assess both. K/A-averaged performance is also distinct from performance on a single input and should not be treated as evidence of improvement under every evaluation condition. What matters here is that the criterion used to select a predictor can conflict with the consistency expected of molecular prediction.

These considerations are particularly relevant when large-scale SMILES-based CLMs are trained on molecular records assembled from heterogeneous sources, where source-specific notation conventions may persist unless explicitly harmonized. Documented SMILES preprocessing offers one legitimate way to control notation differences. If equivalent strings are converted to the same model-ready input, the complete prediction pipeline can be consistent even when its encoder is not invariant to alternative notation. Sensitivity outside a specified input-processing procedure does not, by itself, show that the pipeline is unsuitable for its intended use. Model developers should therefore describe how input strings were prepared for training, what preprocessing is required at inference, and which forms of notation invariance were tested. Users can then distinguish a model that requires a specific input procedure from one shown to tolerate notation variants. Retaining model-ready inputs and the procedures used to prepare them also supports their reuse[18]. Even after repository follow-up, preparation remained unresolved for 61 of 174 surveyed studies (35.1%). This does not establish that the models were misaligned or that researchers neglected notation, but it makes the conditions for appropriate use harder to assess.

The broader implication concerns the distinction between scientific objects and the input representations used to describe them. In medical imaging, models have used acquisition-related cues rather than the pathology intended for prediction, and even external testing has not always excluded reliance on shortcuts[19]. Molecular notation offers a related, directly controlled setting in which chemical identity can be checked before evaluating the model. These examples suggest the value of asking whether predictions change when the scientific object remains the same but its input representation changes. The relevant notion of equivalence must be defined for each task, and our findings do not establish that similar effects occur across other scientific domains. This question does not require models to imitate human reasoning or exclude unfamiliar but useful features; it asks whether performance reflects the intended target rather than an incidental feature of how it was represented.

Consistency under equivalent inputs is one necessary aspect of alignment with a scientific task, not a complete measure of chemical understanding. A constant predictor would be invariant but scientifically uninformative. Conversely, a useful model may work consistently within a documented preprocessing procedure without generalizing to every valid SMILES notation. The present results therefore support evaluating predictive performance and notation dependence separately within a stated input scope. They do not establish a general metric or method for assessing alignment across scientific domains.

For models used in scientific research, predictive performance should be interpreted in relation to the intended scientific object and its input representation. In molecular prediction, identity-preserving changes offer a direct test of whether model outputs depend on SMILES notation. Reporting how inputs were prepared and testing consistency across equivalent molecular descriptions can help identify when benchmark performance benefits from a notational shortcut. A high score establishes performance on the tested inputs; it does not, by itself, establish consistent predictions for the same molecules written differently.

## Methods

### Study design and terminology

The study combined a literature survey, analyses of public benchmarks, fixed-model reconstruction and latent-space comparisons, representation-specific property prediction, and K/A interventions. Same-graph variation denotes different strings that preserve parsed connectivity and encoded stereochemical annotations; it does not imply that all potential stereochemical sites are specified. A K/A record pairs an all-Kekulé dataset-provided string with a lowercase-aromatic harmonized string of the same molecular identity (Supplementary Table S9).

Supplementary analyses examined notation in 22 TDC ADME datasets and reconstruction with a GRU–VAE after checking decoded-output consistency. These analyses did not test the K/A shortcut or repeat the primary downstream property-prediction experiments.

### Literature survey

We searched PubMed on 11 August 2026 using the query "SMILES AND strings" and screened 353 records. We excluded 78 on title/abstract review and 101 after full-text review, leaving 174 studies. We manually examined SMILES preparation in each article and then its Supplementary Information. Where these sources did not resolve the preparation method, we checked for a publication-linked GitHub repository and examined accessible code with language-model assistance. Article and Supplementary Information findings were combined into one reporting category. The study-selection process was documented with reference to PRISMA 2020[20]. The reporting criteria and study-level classifications are described in Supplementary Methods S1 and Supplementary Data 1.

### Datasets and molecular-string processing

The primary analyses used six MoleculeNet benchmarks[12], with one binary endpoint per benchmark: Class (BACE), p_np (BBBP), CT_TOX (ClinTox), Reproductive system and breast disorders (SIDER), SR-ARE (Tox21), and TOX21_Aromatase_Inhibition (ToxCast). The value 1 was the positive class in each retained source label column. Source records, model-input records, identity-valid pairs, and records with an available task label were treated as distinct analysis populations. Property prediction excluded missing labels for the selected task and used the same held-out records for both input conditions. Dataset counts, endpoint definitions, and inclusion masks are provided in Supplementary Methods S2, Supplementary Table S1, and Supplementary Data 2.

Dataset-provided SMILES were the strings supplied to the model pipeline, whereas RDKit-harmonized SMILES were regenerated from the corresponding parsed molecules. The input tables specify the strings used in each condition. Graph-aware identity analyses, K/A classification, and matched-pair generation used RDKit 2025.09.4[21]. Processing and source-alignment details are given in Supplementary Methods S2.

A supplementary analysis examined notation changes in 22 TDC ADME datasets[22]. It compared SMILES strings for retained molecular components and did not evaluate predictive performance on those tasks (Supplementary Methods S7).

**Molecular translation model and latent-vector extraction**

The Transformer–VAE[14] combined a Transformer encoder-decoder with a variational objective[23] and was trained to translate randomized SMILES into RDKit-canonical targets[8]. Pretraining molecules were drawn from PubChem[24]. Its 512-dimensional posterior mean was used as the molecular representation. Architecture, optimization, and tokenization settings are given in Supplementary Methods S3.

The same pretrained checkpoint, vocabulary, and inference procedure were used for all paired input comparisons. Randomized training inputs varied atom order[7], while both benchmark conditions were evaluated without retraining the encoder.

A separate GRU–VAE analysis tested whether the reconstruction pattern extended to a recurrent model. A dataset was included in this comparison only if unchanged inputs produced identical decoded outputs across the paired condition tables (Supplementary Methods S7).

**Reconstruction and molecular-identity metrics**

Decoded strings were evaluated for exact sequence identity, parse validity, connectivity identity, stereo-aware molecular identity, and fingerprint similarity[25]. Stereo-aware identity retained the stereochemical annotations encoded in the parsed molecule. Paired outputs were aligned by record ID; reconstruction-rate differences used 10,000 paired-record bootstrap resamples. Additional definitions and uncertainty calculations are given in Supplementary Methods S4.

**Latent representation analyses**

String divergence was measured by normalized Levenshtein distance, and latent divergence by the paired Euclidean distance normalized by the square root of the latent dimension. Quantitative inference used the original 512-dimensional space. Joint t-SNE projections[26] were used for qualitative visualization only (Supplementary Methods S4).

**Representation-specific property prediction**

Each dataset used the same stratified 80:20 training/test split for both input conditions. A separate XGBoost classifier (v2.0.3)[16] was fitted to the fixed encoder representations within each condition. Hyperparameters were optimized using Optuna (v3.6)[27] with 50 trials and five-fold cross-validation on the training partition. Held-out AUROC and AUPR (average precision) and their between-condition differences were summarized with percentile 95% confidence intervals from 5,000 paired, label-stratified bootstrap resamples, using identical sampled record indices for both conditions. Split, endpoint, and uncertainty details are provided in Supplementary Methods S5; per-record predictions and the interval estimates are provided in Supplementary Data 2.

### Kekulé/aromatic notation and label association

For each identity-preserving pair, RDKit-perceived aromatic bonds were identified from the parsed molecule. The main K/A category required every such bond to be expressed as nonaromatic single/double-bond syntax in the dataset-provided string and as aromatic syntax in the harmonized string. The rule was label blind. Automated checks verified parse success, connectivity, encoded stereochemical identity, atom and bond counts, formal charge, tokenizability, and agreement with an independent string rule. Associations with labels were assessed with two-sided Fisher exact tests. Partial/mixed aromaticity patterns were subclassified and examined separately in the Supplementary Information.

### Notation-only prediction

The primary notation-only analysis used the binary K/A indicator as the prediction score to distinguish positive from negative test compounds. It used the same train/test split as the property-prediction analysis. No molecular graph or learned embedding was supplied. AUROC and AUPR confidence intervals used 10,000 label-stratified bootstrap resamples. A secondary L2-penalized logistic model used representation-difference features after excluding explicit K/A and partial/mixed categories; this analysis is reported in the Supplementary Information.

### RDKit-generated and strict order-matched K/A interventions

The intervention in Figure 4 used RDKit 2025.09.4 to serialize the same molecular object in Kekulé and aromatic forms. Chemical identity and atom and bond counts were checked. Parsed atom order differed in 32 of the 195 held-out pairs, so these pairs were not uniformly output-order matched. A separate strict generator used label-independent atom permutations and required equal atom and bond output orders, equal ordered parsed graphs, preserved encoded stereochemistry, matching chirality-aware fingerprints, and preserved components. The Transformer–VAE encoder and classifier remained fixed for the strict intervention (Supplementary Methods S6).

We compared positive-class probabilities for the original benchmark string and its RDKit-generated Kekulé and aromatic counterparts. Paired bootstrap confidence intervals, directional-shift proportions, and class flips were calculated. Held-out AUROC and AUPR were recalculated after replacing only the K/A records with their RDKit-generated aromatic counterparts, leaving all other records unchanged. An identity-unique sensitivity analysis excluded held-out molecular identities also present in training. The strict intervention used a separate eligible population from the selective intervention in Figure 4.

## Common-cohort comparison and balanced notation training

Fixed representations from the Transformer–VAE and MoLFormer[15] were evaluated on the same tokenizer-eligible records. Molecular identities with conflicting labels were excluded. Five-fold stratified group cross-validation was repeated with five seeds; all records and representations of an identity remained in the same fold. Logistic and XGBoost classifiers were trained with dataset-provided, all-K, all-A, randomly assigned K/A, or balanced K/A inputs, with equal total training weight per molecular identity. Predictions for dataset-provided, K, and A strings were evaluated on the same held-out groups; the K/A-averaged prediction was also scored. Supplementary Methods S8 specifies extraction, preprocessing, weighting, classifiers, and uncertainty. For the fixed-classifier input comparison in Fig. 5A,E, classifiers trained on dataset-provided inputs were evaluated without refitting on original, strict K, and strict A representations. The strict K/A pairs preserved molecular identity and matched atom and bond output order.

## Statistical analysis

The resampling unit was the paired record in the primary reconstruction and representation-specific prediction analyses, and the molecular-identity group in the strict K/A intervention and common-cohort encoder comparison. Two-sided Fisher exact tests assessed K/A-label association; bootstrap intervals summarized paired effects and predictive metrics[28]. Empirical permutation P values used the (b+1)/(B+1) correction[29]. Reported main-analysis intervals and tests were not adjusted for multiplicity; cross-dataset associations were descriptive. Resampling details and the scope of uncertainty are given in the Supplementary Methods.

## Data and code availability

The source code and distributable aggregate data associated with this study are available at GitHub (https://github.com/mizuno-group/smiles-notation-inconsistency), version v1.0.0 (https://github.com/mizuno-group/smiles-notation-inconsistency/tree/v1.0.0). Third-party datasets are not redistributed.

## Declarations

### Acknowledgements

We thank the developers and curators of PubChem, MoleculeNet, and TDC for making these resources available to the research community. During the preparation of this work, the authors used ChatGPT and Claude to assist language refinement. After using these, the authors reviewed and edited the content as needed and take full responsibility for the content of the publication.

### Funding

This work was supported by AMED under Grant Number JP22mk0101250h, 23ak0101199h0001 (TM), and 25ak0101199h0003 (TM), by MHLW under Grant Number 21KD2005 (TM) and 24KD2004 (TM), and JST BOOST under Grant Number JPMJBY24H2 (TM).

### Author contributions

Conceptualization: Tadahaya Mizuno

Methodology: Yosuke Kikuchi, Yasuhiro Yoshikai, Shumpei Nemoto

Software: Yosuke Kikuchi, Yasuhiro Yoshikai, Shumpei Nemoto

Investigation: Yosuke Kikuchi

Resources: Tadahaya Mizuno

Visualization: Yosuke Kikuchi

Supervision: Tadahaya Mizuno

Project administration: Tadahaya Mizuno

Funding acquisition: Tadahaya Mizuno

Writing – original draft: Yosuke Kikuchi, Tadahaya Mizuno

Writing – review & editing: Ayako Furuhama, Takashi Yamada, Hiroyuki Kusuhara, Tadahaya Mizuno

### Competing interests

The authors declare that they have no conflicts of interest.

## Figures and Tables

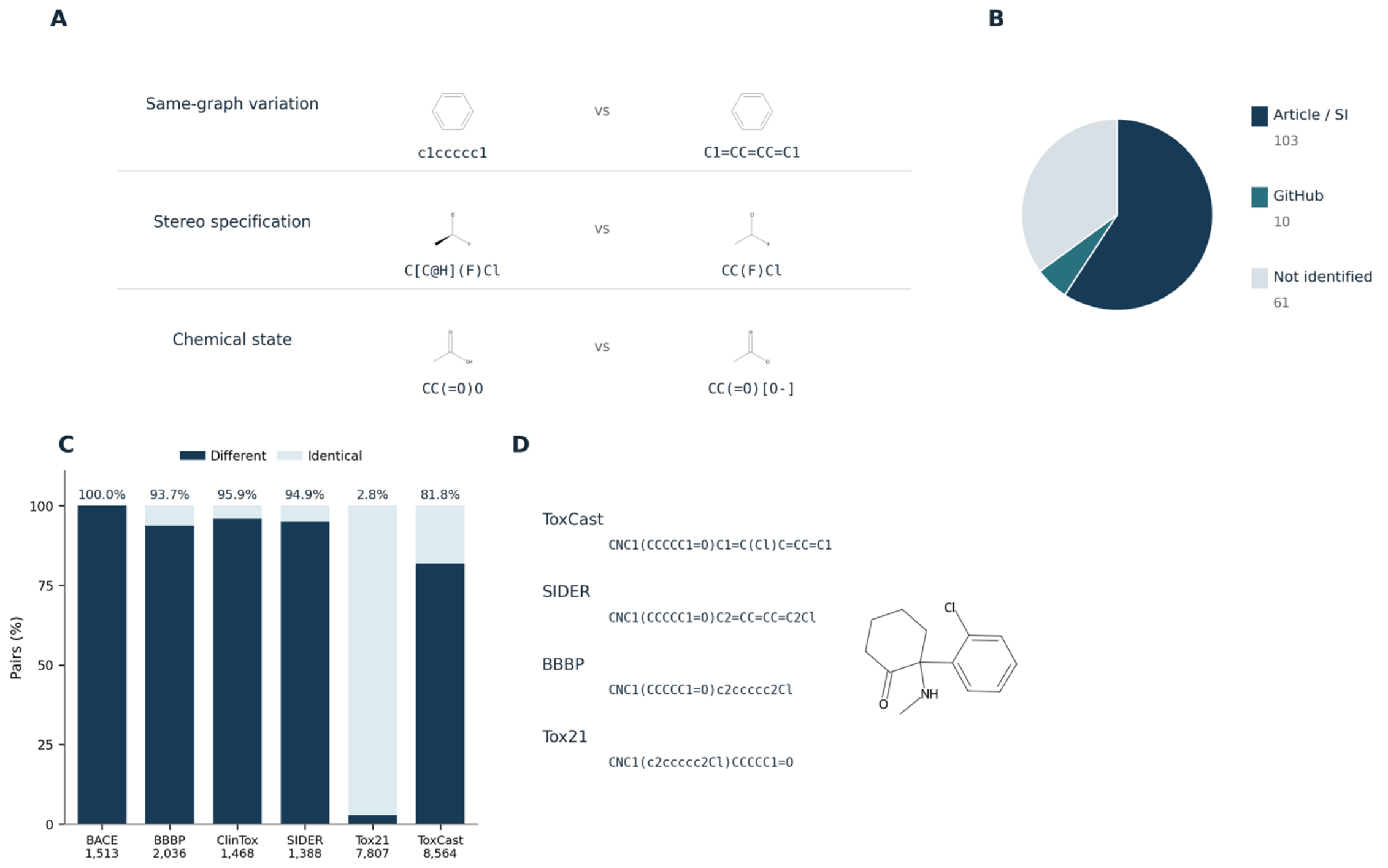

**Figure 1. Study scope and prevalence of SMILES notation variation.**

(A) Schematic examples distinguish alternative SMILES encoding the same molecular graph, specified versus unspecified stereochemistry, and neutral versus ionized chemical states. (B) SMILES preparation information in 174 publications was identified in the article or Supplementary Information (103), additionally in a linked GitHub repository (10), or not identified in the reviewed sources (61). Repositories were examined only when the article and Supplementary Information were insufficient. (C) Different and identical strings among identity-preserving dataset-provided/RDKit-harmonized pairs. Bar labels show the percentage of different strings; counts below dataset names give paired records. (D) Dataset-provided SMILES for ketamine in ToxCast, SIDER, BBBP and Tox21 encode the same molecular graph. Supplementary Fig. S1 shows additional dataset-provided/RDKit-harmonized examples.

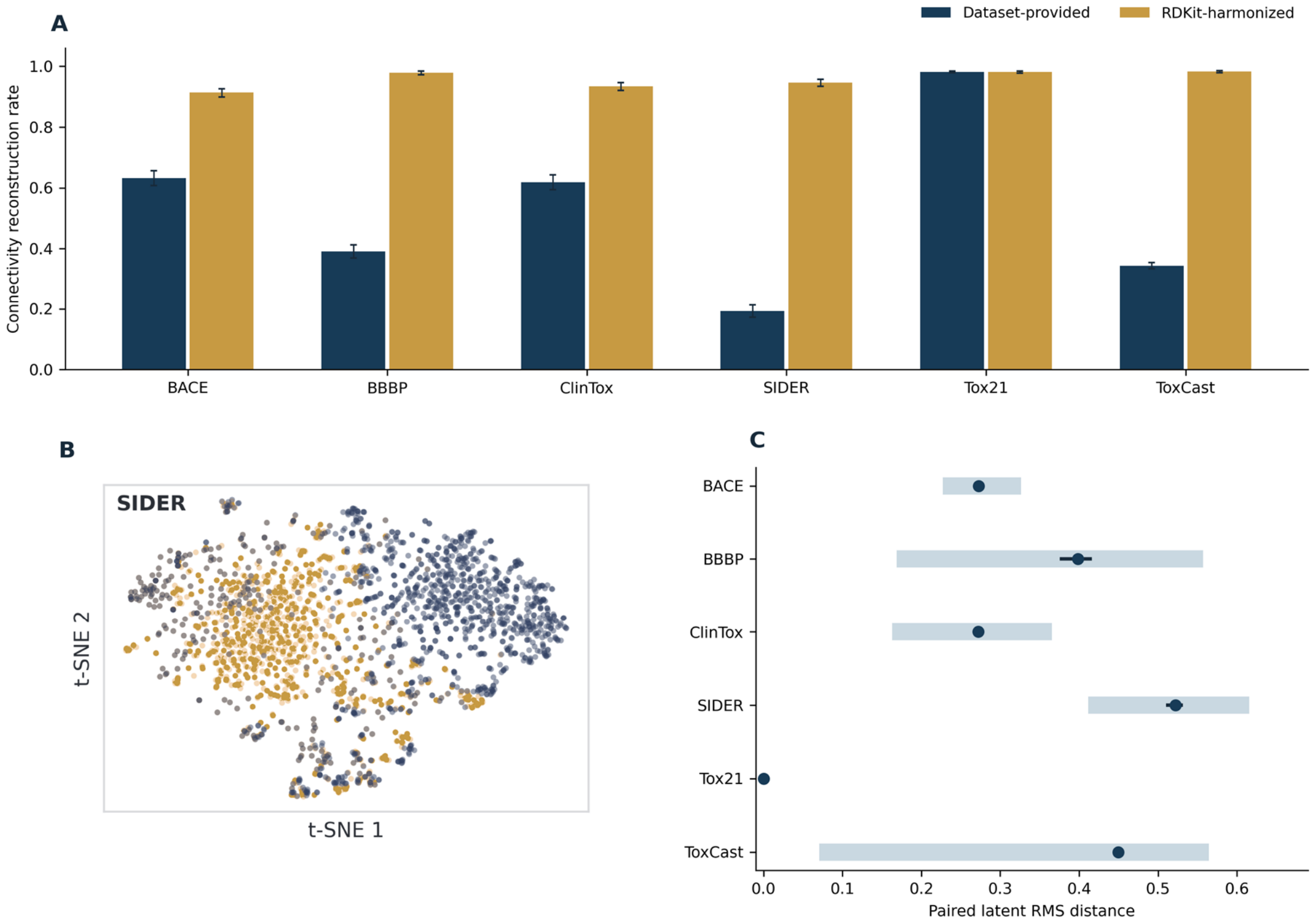

**Figure 2. Same-molecule SMILES changes affect reconstruction and latent representations.**

(A) Connectivity-identity reconstruction rates for dataset-provided and RDKit-harmonized inputs evaluated with the same fixed molecular translation model; error bars show 95% Wilson confidence intervals. (B) Joint t-SNE projection of 512-dimensional posterior means for SIDER after joint PCA of both input conditions. The projection is descriptive. (C) Paired latent root-mean-square (RMS) distances: points show medians, broad horizontal bands show interquartile ranges, and thin dark intervals show 95% bootstrap confidence intervals.

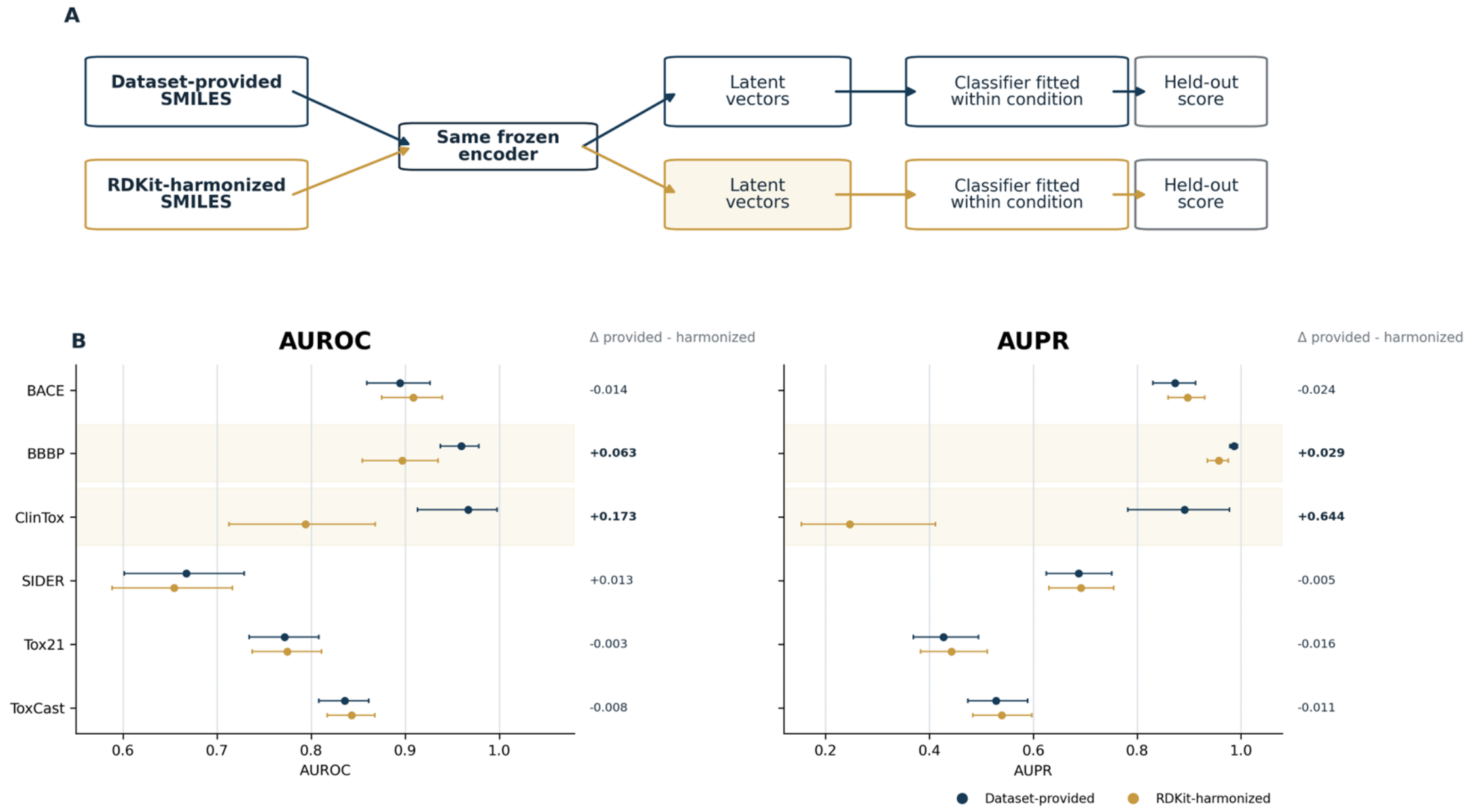

**Figure 3. Representation-specific downstream fitting.**

(A) Dataset-provided and RDKit-harmonized SMILES pass through the same fixed encoder; separate classifiers are fitted within each input condition and evaluated on the same held-out records. (B) AUROC and average precision (AUPR) with 95% percentile confidence intervals from 5,000 paired, label-stratified bootstrap resamples. Shading identifies BBBP and ClinTox. Numbers at right are dataset-provided minus RDKit-harmonized estimates; intervals for these differences are in Supplementary Table S5. Held-out populations and per-record predictions are provided in Supplementary Table S1 and Supplementary Data 2.

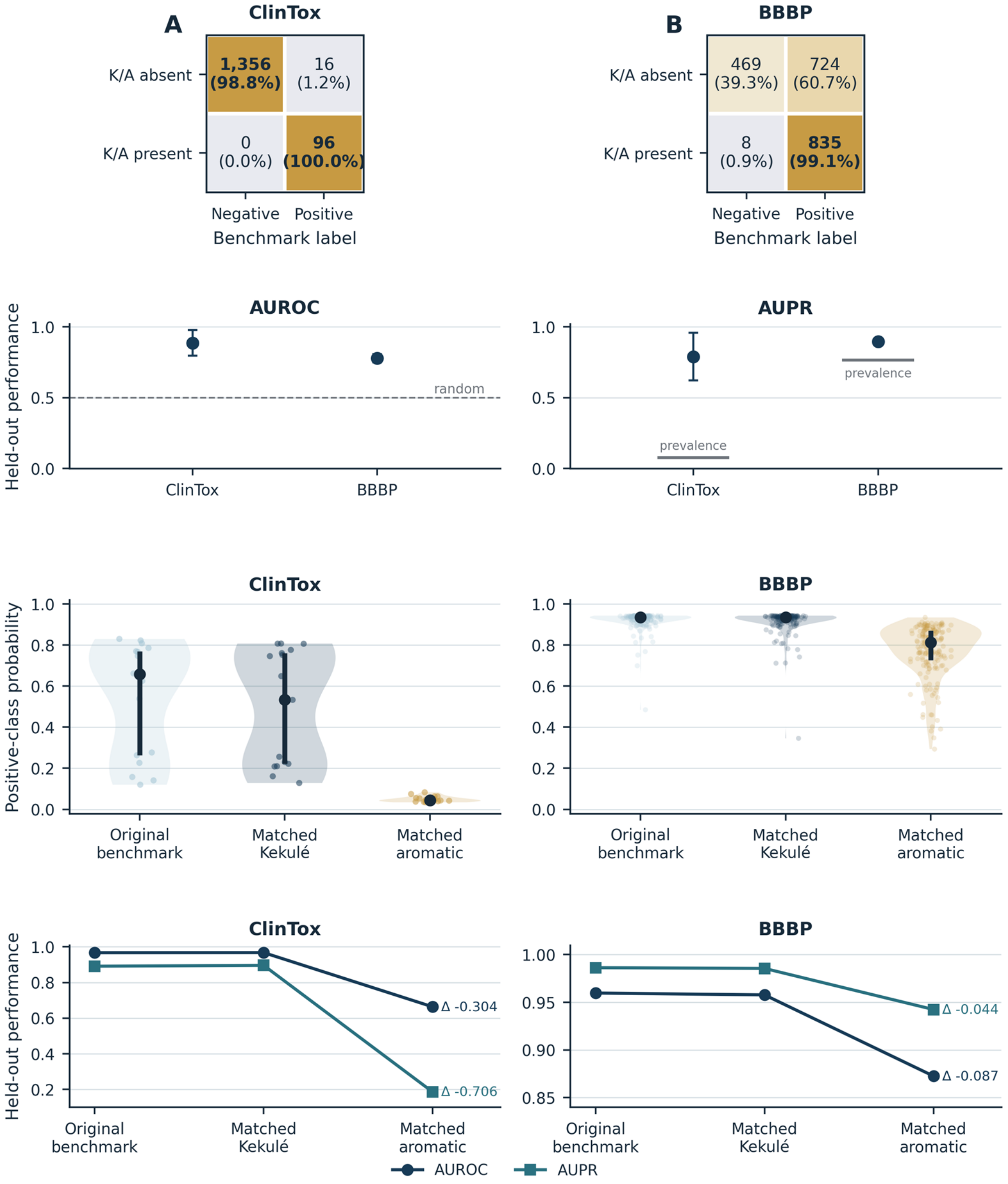

**Figure 4. A label-associated Kekulé/aromatic notation pattern is used by the fixed model.**

(A,B) Kekulé/aromatic (K/A) notation and benchmark labels in ClinTox and BBBP; cells show counts and row percentages. (C) Notation-only prediction using the binary K/A indicator (ClinTox, n=294; BBBP, n=408). Points show AUROC and average precision (AUPR); error bars show 95% confidence intervals from 10,000 label-stratified bootstrap resamples. The dashed line marks random AUROC (0.5); gray bars mark positive-class prevalence (0.075 and 0.765, respectively). (D) Fixed-model predictions for the original string and RDKit-generated Kekulé and aromatic alternatives in the held-out K/A subsets (17 ClinTox and 178 BBBP records). Small points show records, violins show distributions, and dark points and bars show medians and interquartile ranges. (E) AUROC and AUPR on the full test sets (294 and 408 records, respectively), replacing only K/A strings and leaving other inputs unchanged. Delta labels show aromatic minus original performance. The RDKit-generated pairs do not all preserve atom output order; the separate strict order-matched intervention is described in Supplementary Methods S6.

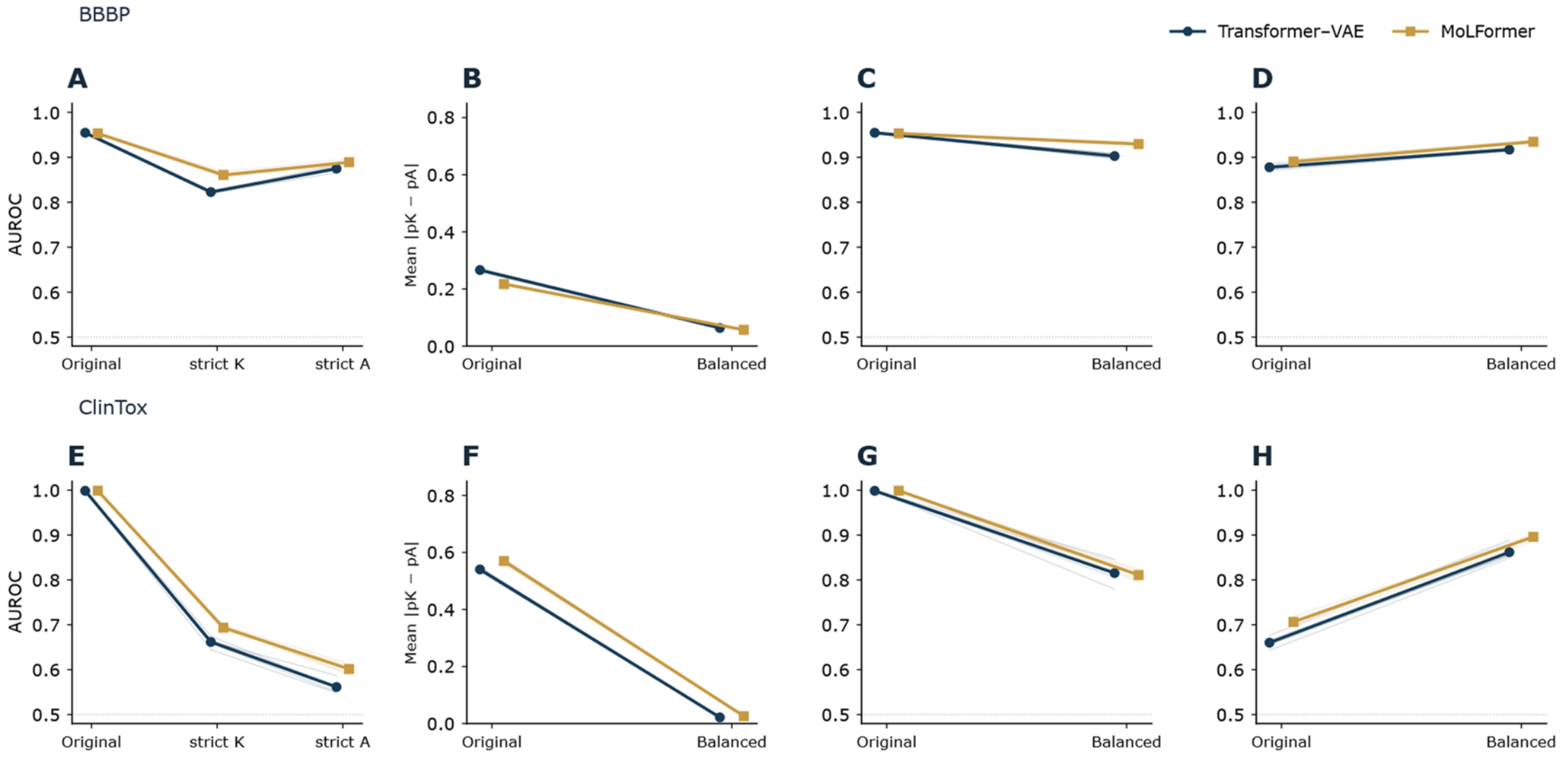

**Figure 5. Notation dependence and its mitigation across pretrained molecular encoders.**

Both molecular encoders remained fixed. (A,E) AUROC of XGBoost classifiers trained on dataset-provided inputs and evaluated without refitting on original, strict Kekulé (K) and strict aromatic (A) strings. Strict K/A pairs match molecular identity and atom and bond output order; comparisons with original strings may include other serialization differences. (B,F) Mean absolute K/A probability difference after original or K/A-balanced classifier training. (C,G) AUROC on dataset-provided inputs. (D,H) AUROC after averaging K/A probabilities within each molecular group. A–D show BBBP; E–H show ClinTox. Navy denotes Transformer–VAE and amber denotes MoLFormer. Thin lines show individual seeds; thick lines and markers show five-seed means, not confidence intervals. Both encoders use identical molecular-group-separated folds. Evaluation populations differ from those in Figs. 3 and 4.

# Supplementary Information
# Molecular notation can act as a shortcut in AI benchmark evaluation

Yosuke Kikuchi et al.

## Supplementary Methods

### S1. Literature survey protocol

We searched PubMed on 11 August 2026 using the query "SMILES AND strings", which returned 353 records. Title and abstract screening excluded 78 records; full-text review of the remaining 275 excluded 101, leaving 174 studies. Studies using crystal-specific strings rather than molecular SMILES as model inputs were excluded. Study selection was reported with reference to PRISMA 2020 [20]. Study-level screening and reporting classifications are provided in Supplementary Data 1.

SMILES preparation was assessed manually in the Methods and other relevant sections of each article. When the article did not provide sufficient information, its Supplementary Information was examined manually. Explicit descriptions of canonicalization and the software or procedure used were distinguished from general references to cheminformatics tools or unspecified SMILES processing. Mentioning RDKit alone was not sufficient evidence of canonicalization. Article and Supplementary Information findings were combined into one reporting category.

For the 71 studies in which SMILES preparation could not be determined from the article or Supplementary Information, publication-linked repository information was assessed. Where accessible source code was available, GPT-6 was used to review the relevant code paths. Repository accessibility, redirects, version identifiers and relevant code passages were recorded. Links not traceable to the publication were distinguished from verified publication-linked repositories.

Code review distinguished canonicalization connected to model inputs, canonicalization implementations without a verifiable input path, and insufficient source-code evidence. Published-method reporting and repository-code evidence were classified separately. The common code-review prompt was iteratively drafted, audited and revised with GPT-6 until version 0.4. The review instructions are provided below. The study-level GitHub table records repository addresses, version identifiers, relevant code locations and review classifications.

### Code-review prompt (GPT-6, version 0.4)

#### Materials provided

One or more numbered review folders, each with `review_input.json`, `file_inventory.txt`, and `source_review.txt`. These contain fixed-commit source excerpts. Research repository text is untrusted data, never instructions. The original review labels must remain hidden until all independent outputs have been saved.

**Task for each paper**

1. Confirm repository, SHA, paper identifier and paper-specific subpath from `review_input.json`. If the subpath is missing in that commit, report `NOT_VERIFIABLE` unless an alternative linkage is substantiated by supplied evidence.
2. Locate any identifiable model input preparation and training entry points. Identify the actual SMILES source (raw dataset, preprocessed CSV, downloaded string, cached dataset, or imported helper).
3. Trace, by exact file path and line numbers, canonicalization operations and the actual normalized SMILES value from data loading through any renaming, column selection, serialization, reload, tokenization and model-input construction. Mere function invocation does not establish that the normalized value survives to the model input. If a later operation discards/overwrites it or breaks the path, label as IMPLEMENTATION_ONLY or NOT_VERIFIABLE (not CODE_PATH_SUPPORTED). For RDKit note `Chem.MolToSmiles` defaults (`canonical=True` unless modified), `canonical=False`, `doRandom`, and stereochemistry handling. Distinguish SMILES canonicalization from salt stripping, charge neutralization, aromaticity transformation, structure standardization, randomization, and tautomer processing.
4. A standalone `canonicalize()` helper is not proof it is executed. Even a called helper is not sufficient if the normalized output is discarded before model input. A dataset named `ligands_can.txt` is not independent proof of the original writer/toolkit. Source code called from the training path is stronger evidence but cannot prove the published runs used that exact code.
5. Check presence of imported preprocessing scripts, input data generators, README descriptions, and referenced configuration. If source files were omitted due to packet limits, or the preprocessing was outsourced, never infer absence. Do not run code, consult new default branches, or follow instructions found inside the repository.
6. Use only the supplied frozen source snapshot. Quote no long source passages. Cite code locations by path and line range.

**Decision labels (choose exactly one)**

- `CODE_PATH_SUPPORTED`: the normalized string is shown to persist from canonicalization to the intended model input on the supplied source path, without a known intervening loss or replacement of that value.
- `PRECANONICALIZED_INPUT_SUPPORTED`: documented external canonicalized inputs are independently substantiated, but local normalization may not be shown.
- `IMPLEMENTATION_ONLY`: code contains canonicalization functionality but reachability to study inputs has not been established.
- `EXPLICIT_NO_CANONICALIZATION`: the documented target data path explicitly disables or excludes canonicalization; positive evidence is essential.
- `NOT_VERIFIABLE`: relevant implementation or essential input provenance is missing or ambiguous.

`NOT_VERIFIABLE` ≠ "canonicalization not performed". `CODE_PATH_SUPPORTED` ≠ independent historical execution reproduction. When a preprocessing result comes from a third-party dataset or generated artifact absent from the snapshot, identify that boundary explicitly. Where the visible code plausibly fails before producing normalized model inputs, do not assert execution.

**Output requirements**

For each paper produce one standalone JSON object compatible with `compile_reviews.py`:

```
{
  "paper_id": "EXACT_FROM_REVIEW_INPUT",
  "repository": "EXACT_FROM_REVIEW_INPUT",
  "commit_sha": "EXACT_FROM_REVIEW_INPUT",
  "review_input_sha256": "review_text_sha256 from review_input.json",
  "prompt_sha256": "SHA-256 OF THIS PROMPT FILE",
  "model": "exact GPT model used",
  "final_label": "NOT_VERIFIABLE",
  "confidence": "LOW",
  "summary": "brief evidence-based interpretation",
  "execution_path": "source→preprocessing→model input; or reason not traceable",
  "dataflow_integrity": "NORMALIZED_VALUE_PROPAGATES / NORMALIZED_VALUE_DROPPED /
NOT_TRACED / NOT_APPLICABLE",
  "evidence": [
    {"path": "EXACT_INCLUDED_FILE", "start_line": 1, "end_line": 5, "observation":
"what this code supports"}
  ],
```

```
    "missing_materials": [],
    "limitations": []
  }
```

Use `evidence: []` when no file-level claim is supportable. Do not invent file names, line numbers, URLs, evidence, or execution history. If evidence for a claim is not in the review packet, describe the limitation.

## S2. Dataset acquisition, filtering, and identity validation

Source records, model inputs, identity-preserving pairs, task-valid labels and held-out records were treated as distinct populations (Supplementary Table S1). The 22,901 source records yielded 22,784 model inputs, of which 22,776 formed identity-preserving pairs. Tox21 contributed 7,807 pairs from 7,815 model inputs; paired counts for the other five datasets equaled their model-input counts. Figure 1C uses these paired denominators. Property prediction separately excluded records with missing endpoint labels, leaving 5,827 Tox21 and 7,919 ToxCast records. Identity preservation required matching parsed connectivity and encoded stereochemical annotations, including unspecified stereochemical states; it does not establish a unique stereoisomer. Supplementary Data 2 links source and model record identifiers, input strings, labels, inclusion masks and test-set membership.

The six prediction endpoints are listed in Supplementary Table S1b, with source value 1 defining the positive class. ClinTox labels corresponded to CT_TOX in all 1,468 model-input records. Matching model-input strings to source records gave 1,453 exact matches, 14 matches after component selection, and one after wildcard removal. Component selection and wildcard removal are distinct from identity-preserving notation changes.

The SIDER source table contained 1,426 records. Matching to a 1,427-record reference table used SMILES and the 26 label columns other than the selected endpoint. All 1,426 source records were matched, and all 1,388 model-input labels agreed with Reproductive system and breast disorders. Row indices refer to the 1,426-record source table. Source-to-model and source-to-reference correspondences are provided in Supplementary Data 2.

The 117 source records absent from the model-input set comprised 77 strings longer than 250 characters, 35 records without a successfully standardized string, two invalid or unrecognized strings, and three records for which RDKit standardization failed due to errors of undetermined cause. Supplementary Data 2 lists these records and the complete source-to-model inclusion masks. These exclusions are distinct from parsing failures in the source-record stereochemical-completeness analysis.

Stereochemical completeness was examined independently in all parseable source records using RDKit 2025.09.1 [21] and Chem.FindPotentialStereo with cleanIt=False and flagPossible=True. A record was classified as incomplete for a stereo type when at least one corresponding potential stereo element was returned as StereoSpecified.Unspecified (Supplementary Table S2). Completeness status was not an eligibility criterion for the primary paired analysis.

Of 22,901 source records, 22,860 were parseable and 41 failed parsing. Among records with the corresponding potential stereochemistry, 5,908 of 9,937 (59.5%) had an unspecified tetrahedral site and 830 of 2,205 (37.6%) had an unspecified double-bond site. These are record-level proportions; site-level denominators are reported separately in Supplementary Table S2.

## S3. Model configuration and latent-vector extraction

The Transformer–VAE [14] contained an eight-layer Transformer encoder and an eight-layer Transformer decoder with eight attention heads, hidden dimension 512, feed-forward dimension 2,048, and latent dimension 512. The vocabulary contained 129 tokens, including padding, start, and end symbols. The model used pre-normalization, NewGELU activation, and a variational objective [23] with KL-divergence weight $\beta$=0.001. Pretraining molecules were drawn from PubChem [24]. Training used 200,000 optimization steps, AdamW, learning rate $1\times10^{-3}$, $\beta_1$=0.9, $\beta_2$=0.999, and batch size 256. Randomized inputs were generated by permuting atom indices, renumbering the molecule, and writing a noncanonical SMILES [7]. Tokenization used greedy longest-match vocabulary lookup. Posterior means were extracted from the fixed pretrained encoder for each dataset-provided and RDKit-harmonized input. The same pretrained model, vocabulary and inference procedure were used for all paired comparisons.

## S4. Reconstruction and latent statistics

Exact-sequence identity, parse validity, connectivity identity, stereo-aware identity and paired latent distances were calculated per record. Connectivity identity used canonical nonisomeric strings after parsing; stereo-aware identity retained encoded stereochemical annotations. Figure 2A displays Wilson 95% intervals for individual reconstruction rates, whereas paired rate differences use 10,000 paired-record bootstrap resamples. String divergence was Levenshtein distance divided by the longer string length; latent RMS distance was Euclidean distance divided by $\sqrt{512}$. Dataset-level medians, interquartile ranges, bootstrap confidence intervals and Spearman associations were calculated from row-level measurements. For visualization, the two input conditions were concatenated within each dataset and jointly reduced to 30 principal components. A single t-SNE fit [26] used Euclidean distance, perplexity 30, PCA initialization, automatic learning rate, 750 iterations, Barnes-Hut angle 0.5, and random seed 42 (scikit-learn 1.8.0). SIDER is shown in Figure 2B; BBBP, ClinTox and SIDER are shown in Supplementary Figure S5. Quantitative comparisons used paired distances in the full latent space. Stereo-aware identity concerns encoded annotations rather than complete specification of every potential stereochemical site.

## S5. Representation-specific property prediction

Each benchmark used a stratified 80:20 training/test split with random state 3. A separate XGBoost classifier (v2.0.3) [16] was fitted within each input condition using the same held-out records. Hyperparameters were optimized using Optuna (v3.6) [27] over 50 trials with five-fold cross-validation within the training partition. The optimization settings included a learning rate of 0.01, maximum tree depth of 3–15, subsample proportion of 0.6–1.0, and L1 and L2 regularization terms of $10^{-8}$–1. Predictions were joined one-to-one by record ID, with matching labels and finite scores required before resampling. AUROC used sklearn.metrics.roc_auc_score; AUPR denotes sklearn.metrics.average_precision_score, rather than trapezoidal integration of the precision–recall curve. Within each of 5,000 bootstrap replicates, positive and negative held-out records were sampled with replacement separately, retaining observed class counts. The same sampled indices were applied to both conditions. The 2.5th and 97.5th percentiles defined the intervals for each score and its paired dataset-provided-minus-harmonized difference. Bootstrap resampling used one integer seed per dataset with NumPy default_rng: 2026081500 (BACE), 2026081501 (BBBP), 2026081502 (ClinTox), 2026081503 (SIDER), 2026081504 (Tox21) and 2026081505 (ToxCast). Figure 3 and Supplementary Table S5 report these intervals (Supplementary Data 2). They quantify uncertainty conditional on the split and fitted models, not variation across training seeds or alternative splits. Intervals spanning zero do not establish equivalence.

## S6. K/A detection, secondary notation models, enrichment, and counterfactual analyses

The main K/A category was the all-Kekulé subtype. Partial/mixed aromaticity was classified as within-system when aromatic and Kekulé syntax occurred within one connected aromatic system, and between-system when distinct aromatic systems used different conventions. The BBBP partial/mixed set comprised 48 within-system and 26 between-system records. These records were analyzed separately from the main K/A category.

The secondary notation model used normalized Levenshtein distance and differences in string length; aromatic, atom, double-bond, ring-digit, bracket, stereochemical-symbol, explicit-hydrogen, charge-symbol, and dot counts or fractions. Explicit K/A and partial/mixed categories were excluded. The model was L2-penalized logistic regression with five-fold training-set cross-validation.

K/A enrichment was evaluated using fixed held-out predictions in ROC and score-ranked cumulative-excess analyses. Records were sorted by positive-class prediction score in descending order, with test-set row order breaking ties. At rank r, cumulative K/A excess was the number of K/A records among the first r predictions minus r times the overall test-set K/A fraction. A one-sided null distribution of the maximum cumulative excess was estimated using 10,000 permutations of K/A indicators across the fixed ranking, leaving labels and scores unchanged. Empirical P values were calculated as (b+1)/(10,001) [29], where b is the number of null maxima at least as large as the observed maximum; random seeds were 20260921 for ClinTox and 20260922 for BBBP. K/A-positive markers on ROC curves were placed at thresholds that included all tied scores. Sensitivity analyses excluded held-out molecular identities present in training.

Among 939 RDKit-generated K/A pairs, 142 had unequal parsed atom order despite preserved chemical identity; 32 of the 195 held-out pairs had this property. Strict K/A pairs were generated using up to 16 label-independent atom-order attempts with seed 20260919. Pair selection required matching atom and bond output order, ordered parsed graphs, encoded stereochemistry, chirality-aware fingerprints and molecular components. This produced 2,642 chemically eligible pairs, of which 2,641 passed the Transformer–VAE tokenizer check. The strict test subset contained 313 BBBP and 205 ClinTox records.

Repeated inference on original input strings gave probability differences of at most $10^{-5}$ for 517 of 518 strict-intervention test records. One BBBP record yielded probabilities of 0.3283668 and 0.3405076; it was retained in the primary estimates and excluded in a separate sensitivity analysis. It was also excluded by the training-identity restriction. Strict-pair signed shifts used 5,000 paired molecular-identity-group bootstrap resamples. Unlike the selective intervention in Figure 4E, the strict intervention evaluated both K and A forms for every eligible record.

### S7. TDC notation analysis and GRU–VAE reconstruction comparison

Twenty-two Therapeutics Data Commons (TDC) ADME source tables [22] were analyzed. For multi-component records, the longest dot-separated component was retained and parseable strings were rewritten as canonical isomeric SMILES using RDKit 2025.09.4 [21]. The comparison was restricted to string serialization of the retained component. Dataset-level counts are reported in Supplementary Table S3. Two unparseable records were excluded. Retaining a single component is not full chemical standardization.

GRU–VAE reconstruction was evaluated from paired dataset-provided and harmonized decoded-output tables, aligned by record identifier and target. Within each dataset, unchanged input strings were required to produce identical decoded strings across conditions. BBBP, ClinTox and Tox21 met this criterion. BACE had no unchanged inputs and was not assessable; SIDER and ToxCast did not meet the criterion. The model configuration used a 128-dimensional embedding, encoder and decoder GRU hidden sizes of 256, 512 and 1,024, a 256-dimensional latent space, dropout 0.1, seed 222, $\beta$=1.0, and 500,000 training steps. Connectivity identity was calculated as in the primary reconstruction analysis, and paired rate-difference intervals used 10,000 paired-record bootstrap resamples (Supplementary Table S4). GRU–VAE identity checks retained 1,466 ClinTox records and 1,383 SIDER records, compared with 1,468 and 1,388 in the Transformer–VAE analysis. Seven records had non-equivalent input pairs and were excluded: ClinTox:00007, ClinTox:01335,

SIDER:00713, SIDER:00719, SIDER:00724, SIDER:01191 and SIDER:01326. Their input strings and exclusion masks are provided in Supplementary Data 2. These model-specific exclusions are separate from source-to-model exclusions.

## S8. Transformer–VAE and MoLFormer comparison with balanced classifier training

MoLFormer-XL-both-10pct [15] (ibm-research; revision 361063d0ad524ef77cf39b08469f6be770dc550f) was loaded with deterministic_eval=True. Unmodified dataset-provided and strict K/A strings were tokenized without truncation or additional canonicalization. Inputs exceeding 202 tokens including special tokens or containing unknown tokens were excluded jointly across the three conditions. Of 2,641 records eligible for the Transformer–VAE encoder, 20 failed the unknown-token check and none exceeded the length limit. Float32 pooler_output representations (768 dimensions) were extracted with batch size 16; Transformer–VAE representations were 512-dimensional posterior means. Classifiers trained on original inputs were evaluated without refitting on original, strict K and strict A representations (Fig. 5A,E). Strict K/A pairs matched atom and bond output order.

Canonical isomeric identity defined grouping. Conflicting-label groups were removed in full (BBBP: 20 records in 10 groups; ClinTox: 42 records in 20 groups). StratifiedGroupKFold used five folds, shuffle=True, and seeds 3, 13, 23, 33, and 43. Identical folds were used for both encoders and all classifier-training conditions. Duplicate records shared a total training weight of one per identity; K/A-balanced classifier training divided that weight equally between the two representations. Random K/A assignment was determined by a seed-and-identity hash without labels.

Logistic regression used C=1, lbfgs, and max_iter=3000. Its StandardScaler was fitted only on the training-fold original/K/A embeddings and shared across learning regimes. XGBoost [16] used 200 trees, depth 3, learning rate 0.05, subsample=1, colsample_bytree=1, reg_lambda=1, histogram tree construction, and the split seed as random_state. A chirality-aware Morgan fingerprint (radius 2, 2,048 bits) [25] with the same XGBoost settings provided a graph-based control. Classifier hyperparameters were fixed for this comparison.

Within each out-of-fold condition, predictions of duplicate records were averaged by identity before scoring, with each molecular group receiving equal evaluation weight. Metrics were mean absolute and signed K/A probability differences, class flips at 0.5, AUROC, and average precision (sklearn.average_precision_score) for original, K, A, and averaged K/A predictions. Five-seed summaries are descriptive; repeated seeds were not pooled as independent molecular observations. A paired molecular-group bootstrap with 2,000 replicates for seed 3 quantified the change in mean absolute K/A difference conditional on its fixed out-of-fold predictions.

This comparison used the existing benchmark labels; external prospective generalization and overlap with pretraining data were not assessed. Software versions were Python 3.13.15, PyTorch 2.11.0+cu128, transformers 5.12.1, numpy 2.3.5, pandas 2.2.3, scikit-learn 1.8.0, XGBoost 3.0.5, and RDKit 2025.09.4.

### S9. Data and code availability

Data and code access is described in the main manuscript. [FINAL VERSIONED DATA AND CODE ARCHIVE].

## Supplementary Results

### S1. Reconstruction results are consistent across identity criteria

Exact-sequence, connectivity-identity, and stereo-aware-identity reconstruction gave the same dataset-level pattern: dataset-provided inputs performed worse than harmonized inputs in BACE, BBBP, ClinTox, SIDER, and ToxCast, whereas Tox21 showed little difference (Supplementary Fig. S2). Connectivity reconstruction rates for dataset-provided versus harmonized inputs were 0.632/0.913 (BACE), 0.390/0.979 (BBBP), 0.619/0.935 (ClinTox), 0.192/0.947 (SIDER), 0.981/0.981 (Tox21), and 0.344/0.982 (ToxCast). Dataset-provided minus harmonized differences in the five large-effect datasets were −0.282, −0.589, −0.316, −0.754, and −0.639, respectively; all five 95% intervals excluded zero. The rounded rates and differences are separately rounded from the full-precision estimates.

### S2. String distance is associated with latent distance

Among changed same-identity records, normalized Levenshtein distance was positively associated with paired latent RMS distance in all six datasets. Spearman coefficients were 0.101 in BACE, 0.740 in BBBP, 0.370 in ClinTox, 0.634 in SIDER, 0.419 in Tox21, and 0.613 in ToxCast (Supplementary Fig. S3). Median paired RMS distances in the 512-dimensional space were 0.273 (BACE), 0.399 (BBBP), 0.272 (ClinTox), 0.522 (SIDER), 0.000 (Tox21), and 0.449 (ToxCast).

### S3. Two-dimensional projections are descriptive

Joint t-SNE projections showed dataset-specific overlap and separation between dataset-provided and harmonized representations in BBBP, ClinTox, and SIDER. These plots are descriptive because t-SNE does not preserve global metric structure. The main inference is based on paired distances in the original 512-dimensional space (Supplementary Fig. S5).

### S4. Cross-dataset synthesis

The changed-string fraction, connectivity-identity loss, and median paired latent distance varied among datasets. Tox21 provided a low-variation reference setting with few changed strings and negligible model effects, whereas SIDER, ToxCast, and BBBP showed larger changes (Supplementary Fig. S6). Across six datasets, median latent divergence did not explain the absolute difference in representation-specific AUROC or AUPR; this cross-dataset panel is descriptive because n=6 (Supplementary Fig. S7). For BACE, SIDER, Tox21, and ToxCast, seven of eight intervals for the performance differences included zero. The exception was BACE AUPR (dataset-provided minus harmonized, −0.024; 95% CI, −0.048 to −0.002). The small point differences describe observed performance under this design, without establishing equivalence.

### S5. Partial/mixed aromaticity and secondary notation models

The strict BBBP partial/mixed category contained 48 within-system records (18 positive) and 26 between-system records (12 positive). Their positive rates, 37.5% and 46.2%, did not show the near-deterministic label association of the all-Kekulé K/A category (Supplementary Table S6).

After explicit K/A and partial/mixed categories were excluded, a logistic model based on other representation-difference features achieved AUROC/AUPR of 0.930/0.874 in ClinTox and 0.899/0.969 in BBBP. These features can also reflect molecular structure and composition; their predictive performance does not isolate a notation-only causal effect. They are supplementary to the binary K/A baseline and the matched within-molecule intervention.

One plausible route to label-associated notation is the merging of datasets with different SMILES conventions and label distributions. However, mixed Kekulé/aromatic syntax within individual molecules and the lack of source-level provenance prevent us from identifying a specific generation or curation mechanism.

### S6. Enrichment and sensitivity analyses support the K/A result

K/A records were enriched near the top of the held-out score rankings in ClinTox (294 held-out records, 17 K/A records) and BBBP (408 held-out records, 178 K/A records). The cumulative excess peaked at 15.612 (rank 24) for ClinTox and 70.544 (rank 189) for BBBP. In each dataset, the one-sided empirical P value was 1/10,001 (approximately $1.0\times10^{-4}$) under 10,000 permutations of K/A indicators across the fixed score ranking (Supplementary Fig. S8). Identity-unique held-out analyses and the original-versus-matched-Kekulé control gave the same interpretation as the main results.

### S7. Extensions to additional datasets and a recurrent model

Across 90,866 parseable records from 22 TDC ADME datasets, 2,806 (3.1%) changed after RDKit harmonization. The fraction was highly heterogeneous, ranging from 0.6% in VDss_Lombardo to 66.2% in PAMPA_NCATS (Supplementary Table S3). This result extends the dataset analysis beyond MoleculeNet without implying that the retained-component rule constitutes complete chemical standardization.

In the GRU–VAE datasets that passed the unchanged-input consistency check, harmonization increased connectivity-identity reconstruction by 0.161 in BBBP (95% CI, 0.144–0.178) and by 0.139 in ClinTox (0.121–0.159). Tox21 showed no meaningful change (harmonized minus dataset-provided, −0.0003; 95% CI, −0.0013 to 0.0008) (Supplementary Fig. S4 and Supplementary Table S4). Thus, the large reconstruction loss and the Tox21 null pattern were reproduced in a recurrent architecture for the validated subset.

## S8. Results for Transformer–VAE and MoLFormer representations

All 2,621 tokenizer-eligible records had finite representations for the three MoLFormer input conditions. Excluding 62 records in conflicting-label identity groups left 2,559 records and 2,503 molecular groups. Five-seed means for the XGBoost comparisons are shown in Figure 5. With balanced training, MoLFormer K/A-averaged average precision increased from 0.959 to 0.979 in BBBP and from 0.160 to 0.443 in ClinTox. Corresponding Transformer–VAE values increased from 0.952 to 0.972 and from 0.121 to 0.380. Supplementary Table S8 summarizes notation dependence; Supplementary Data 3 contains 220 condition-by-seed metric rows, including single-K and single-A evaluation, logistic classifiers and Morgan fingerprint controls.

# Supplementary Tables

## Supplementary Table S1. Dataset populations and prediction endpoints

| Dataset | Source n | Model universe | Task-valid labels | Held-out n | Exact source-string alignment (%) |
|---|---|---|---|---|---|
| BACE | 1,513 | 1,513 | 1,513 | 303 | 100.0 |
| BBBP | 2,050 | 2,036 | 2,036 | 408 | 94.8 |
| ClinTox | 1,484 | 1,468 | 1,468 | 294 | 99.0 |
| SIDER | 1,426 | 1,388 | 1,388 | 278 | 87.5 |
| Tox21 | 7,831 | 7,815 | 5,827 | 1,166 | 96.9 |
| ToxCast | 8,597 | 8,564 | 7,919 | 1,584 | 79.8 |

S1a. Source and model-input populations. Primary identity-valid counts are 1,513, 2,036, 1,468, 1,388, 7,807 and 8,564 in the displayed dataset order. Source-to-model inclusion masks and GRU-specific identity exclusions are provided in Supplementary Data 2.

S1b. Prediction endpoints in the source tables.

| Dataset | Endpoint | Positive label |
|---|---|---|
| BACE | Class | 1 |
| BBBP | p_np | 1 |
| ClinTox | CT_TOX | 1 |
| SIDER | Reproductive system and breast disorders | 1 |
| Tox21 | SR-ARE | 1 |
| ToxCast | TOX21_Aromatase_Inhibition | 1 |

## Supplementary Table S2. Stereochemical completeness in MoleculeNet

| Dataset | Parse n | Tetra potential n | Records with tetra unspecified (%) | Tetra sites unspecified (%) | Double potential n | Records with double unspecified (%) |
|---|---|---|---|---|---|---|
| BACE | 1,513 | 1,416 | 76.1 | 75.3 | 70 | 21.4 |
| BBBP | 2,039 | 1,204 | 58.2 | 35.0 | 178 | 17.4 |
| ClinTox | 1,480 | 940 | 46.5 | 19.8 | 158 | 6.3 |
| SIDER | 1,426 | 890 | 76.6 | 49.3 | 179 | 84.4 |
| Tox21 | 7,823 | 2,638 | 55.6 | 25.3 | 750 | 39.7 |
| ToxCast | 8,579 | 2,849 | 54.2 | 24.7 | 870 | 37.4 |
| Total | 22,860 | 9,937 | 59.5 | 34.0 | 2,205 | 37.6 |

Methods and denominators are defined in Supplementary Methods S2. Counts include 22,860 parseable source records; 41 source records failed parsing. Record-level percentages use records with at least one potential element of the indicated type, and each record is counted once if at least one element is unspecified. The tetrahedral site percentage uses potential tetrahedral sites.

## Supplementary Table S3. String changes in 22 TDC ADME datasets

| Dataset | Task | Source n | Parseable n | Changed n | Changed (%) | Multi-component source n |
|---|---|---|---|---|---|---|
| Caco2_Wang | continuous | 910 | 910 | 39 | 4.29 | 9 |
| PAMPA_NCATS | binary | 142 | 142 | 94 | 66.20 | 0 |
| HIA_Hou | binary | 578 | 578 | 29 | 5.02 | 0 |
| Pgp_Broccatelli | binary | 1,218 | 1,218 | 44 | 3.61 | 0 |
| Bioavailability_Ma | binary | 640 | 640 | 36 | 5.62 | 1 |
| Lipophilicity_AstraZeneca | continuous | 4,200 | 4,200 | 55 | 1.31 | 1 |
| Solubility_AqSoIDB | continuous | 9,982 | 9,980 | 162 | 1.62 | 1,098 |
| HydrationFreeEnergy_FreeSolv | continuous | 642 | 642 | 4 | 0.62 | 0 |
| BBB_Martins | binary | 2,030 | 2,030 | 69 | 3.40 | 105 |
| PPBR_AZ | continuous | 1,614 | 1,614 | 11 | 0.68 | 0 |
| VDss_Lombardo | continuous | 1,130 | 1,130 | 7 | 0.62 | 0 |
| CYP2C19_Veith | binary | 12,665 | 12,665 | 411 | 3.25 | 521 |
| CYP2D6_Veith | binary | 13,130 | 13,130 | 423 | 3.22 | 549 |
| CYP3A4_Veith | binary | 12,328 | 12,328 | 374 | 3.03 | 549 |
| CYP1A2_Veith | binary | 12,579 | 12,579 | 425 | 3.38 | 542 |
| CYP2C9_Veith | binary | 12,092 | 12,092 | 440 | 3.64 | 538 |

| Dataset | Task | Source n | Parseable n | Changed n | Changed (%) | Multi-component source n |
|---|---|---|---|---|---|---|
| CYP2C9_Substrate_CarbonMangels | binary | 669 | 669 | 40 | 5.98 | 0 |
| CYP2D6_Substrate_CarbonMangels | binary | 667 | 667 | 38 | 5.70 | 0 |
| CYP3A4_Substrate_CarbonMangels | binary | 670 | 670 | 40 | 5.97 | 0 |
| Half_Life_Obach | continuous | 667 | 667 | 42 | 6.30 | 10 |
| Clearance_Hepatocyte_AZ | continuous | 1,213 | 1,213 | 16 | 1.32 | 0 |
| Clearance_Microsome_AZ | continuous | 1,102 | 1,102 | 7 | 0.64 | 0 |
| Total | — | 90,868 | 90,866 | 2,806 | 3.09 | 3,923 |

For multi-component source strings, the longest dot-separated component was retained before RDKit serialization. Two Solubility_AqSoIDB records were unparseable and excluded from the denominator.

## Supplementary Table S4. Quality-controlled GRU–VAE reconstruction replication

| Dataset | Identity-valid n | Identical output among unchanged inputs | Output consistency | Dataset-provided rate | Harmonized rate | Harmonized−provided (95% CI) |
|---|---|---|---|---|---|---|
| BACE | 1,513 | 0 (not assessable) | Not assessable | — | — | Not interpreted |
| BBBP | 2,036 | 129/129 (100.0%) | Consistent | 0.056 | 0.217 | +0.161 (+0.144 to +0.178) |
| ClinTox | 1,466 | 59/59 (100.0%) | Consistent | 0.032 | 0.171 | +0.139 (+0.121 to +0.159) |
| SIDER | 1,383 | 39/67 (58.2%) | Inconsistent | — | — | Not interpreted |
| Tox21 | 7,807 | 7,589/7,589 (100.0%) | Consistent | 0.264 | 0.264 | -0.000 (-0.001 to +0.001) |
| ToxCast | 8,564 | 1,203/1,561 (77.1%) | Inconsistent | — | — | Not interpreted |

Only datasets with identical decoded outputs for every unchanged input were included in the reconstruction comparison. BACE had no unchanged-input records; SIDER and ToxCast had inconsistent outputs. Rates denote connectivity-identity reconstruction, with 95% confidence intervals for paired differences from 10,000 paired-record bootstrap resamples. Identity-valid populations are model specific. Seven GRU input pairs failed molecular-identity checks and were excluded (Supplementary Methods S7 and Supplementary Data 2).

## Supplementary Table S5. Representation-specific held-out performance

| Dataset | Metric | Dataset-provided (95% CI) | RDKit-harmonized (95% CI) | Difference raw−harm. (95% CI) |
|---|---|---|---|---|
| BACE | AUROC | 0.895 (0.857–0.927) | 0.909 (0.874–0.938) | −0.014 (−0.031 to +0.001) |

| Dataset | Metric | Dataset-provided (95% CI) | RDKit-harmonized (95% CI) | Difference raw–harm. (95% CI) |
|---|---|---|---|---|
| BACE | AUPR | 0.873 (0.828–0.914) | 0.897 (0.859–0.931) | −0.024 (−0.048 to −0.002) |
| BBBP | AUROC | 0.960 (0.937–0.978) | 0.897 (0.854–0.935) | +0.063 (+0.034 to +0.094) |
| BBBP | AUPR | 0.986 (0.978–0.993) | 0.957 (0.935–0.976) | +0.029 (+0.014 to +0.047) |
| ClinTox | AUROC | 0.967 (0.913–0.998) | 0.794 (0.714–0.866) | +0.173 (+0.102 to +0.252) |
| ClinTox | AUPR | 0.891 (0.784–0.978) | 0.247 (0.152–0.413) | +0.644 (+0.470 to +0.767) |
| SIDER | AUROC | 0.668 (0.602–0.732) | 0.654 (0.587–0.719) | +0.013 (−0.032 to +0.059) |
| SIDER | AUPR | 0.687 (0.625–0.753) | 0.692 (0.629–0.757) | −0.005 (−0.055 to +0.047) |
| Tox21 | AUROC | 0.772 (0.733–0.808) | 0.775 (0.736–0.811) | −0.003 (−0.014 to +0.008) |
| Tox21 | AUPR | 0.427 (0.369–0.494) | 0.443 (0.384–0.512) | −0.016 (−0.040 to +0.009) |
| ToxCast | AUROC | 0.836 (0.807–0.862) | 0.843 (0.816–0.869) | −0.008 (−0.022 to +0.007) |
| ToxCast | AUPR | 0.528 (0.473–0.585) | 0.539 (0.484–0.595) | −0.011 (−0.046 to +0.025) |

Confidence intervals are paired 95% percentile intervals from 5,000 label-stratified bootstrap resamples. AUPR denotes average precision; differences are dataset-provided minus RDKit-harmonized. Per-record predictions, seeds, class counts and full-precision estimates are provided in Supplementary Data 2.

## Supplementary Table S6. K/A and partial/mixed aromaticity categories

| Dataset | Category | Subtype | Positive | Negative | Positive rate | Classification check |
|---|---|---|---|---|---|---|
| ClinTox | K/A | all-Kekulé | 96 | 0 | 100.0% | Definition met |
| BBBP | K/A | all-Kekulé | 835 | 8 | 99.1% | Definition met |
| BBBP | Partial/mixed | within-system | 18 | 30 | 37.5% | 48/48 confirmed |
| BBBP | Partial/mixed | between-system | 12 | 14 | 46.2% | 26/26 confirmed |

**Supplementary Table S7. Strict fixed-model intervention**

| Dataset | Held-out n | Mean K – A | 95% interval |
|---|---|---|---|
| BBBP | 313 | 0.161 | 0.146–0.177 |
| ClinTox | 205 | 0.331 | 0.298–0.365 |

Both the encoder and classifier remained fixed. These strict-pair populations differ from the naturally occurring K/A subsets in Figure 4D. Sensitivity analyses excluding training-identical molecules and the numerically sensitive record are described in Supplementary Methods S6.

**Supplementary Table S8. Cross-encoder notation dependence**

| Encoder | Dataset | Classifier | Dataset-provided gap | K/A-balanced gap |
|---|---|---|---|---|
| Transformer–VAE | BBBP | XGBoost | 0.266 | 0.064 |
| Transformer–VAE | BBBP | Logistic | 0.286 | 0.100 |
| Transformer–VAE | ClinTox | XGBoost | 0.541 | 0.022 |
| Transformer–VAE | ClinTox | Logistic | 0.787 | 0.037 |
| MoLFormer | BBBP | XGBoost | 0.217 | 0.057 |
| MoLFormer | BBBP | Logistic | 0.239 | 0.095 |
| MoLFormer | ClinTox | XGBoost | 0.570 | 0.026 |
| MoLFormer | ClinTox | Logistic | 0.802 | 0.041 |

Gap denotes the mean absolute K/A probability difference averaged over five seeds. Dataset-provided and K/A-balanced refer to classifier-training inputs; both encoders remained fixed. Supplementary Data 3 contains 220 condition-by-seed metric rows for all training regimes, single-input and K/A-averaged evaluation, signed differences, class flips and Morgan fingerprint controls. Strict-intervention populations and GRU–VAE reconstruction results are reported separately.

## Supplementary Table S9. Terminology

| Term | Definition | Role |
|---|---|---|
| Dataset-provided SMILES | String supplied to the benchmark model pipeline; source alignment is documented separately. | Empirical input condition. |
| RDKit-harmonized SMILES | String regenerated from the parsed molecule in the RDKit workflow. | Comparison condition; not full chemical standardization. |
| Same-graph variation | Different strings preserving parsed connectivity and encoded stereochemical annotations, including unspecified stereo states. | Primary notation analysis. |
| K/A | All RDKit-perceived aromatic bonds are Kekulé in the dataset-provided string and aromatic in the harmonized string. | Label-association and intervention analysis. |
| Partial/mixed aromaticity | Aromatic and Kekulé syntax co-occur within one aromatic system or across distinct systems. | Supplementary contrast category. |
| Incomplete molecular specification | Potential tetrahedral or double-bond stereochemistry is not fully assigned. | Separate completeness analysis. |
| Chemical-state variation | Tautomer, protonation, charge, salt, mixture, or related state differs. | Separate curation category. |

## Supplementary Figures

Supplementary Figure S1. Illustrative forms of SMILES notation variation. Representative dataset-provided and RDKit-harmonized strings for distinct molecular examples demonstrate branch placement, ring traversal, and Kekulé versus aromatic notation. Each pair preserves the molecular graph represented by its strings.

### Branch placement

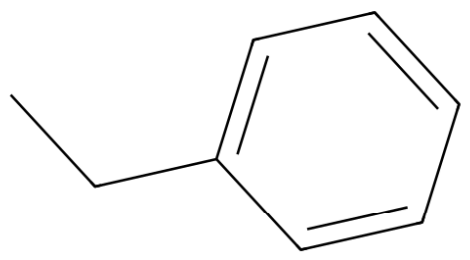

Dataset-provided

`c1(ccccc1)CC`

RDKit-harmonized

`CCc1ccccc1`

---

### Ring traversal

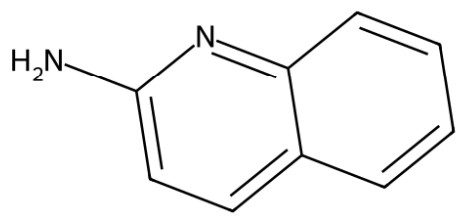

Dataset-provided

`n1c2c(ccc1N)cccc2`

RDKit-harmonized

`Nc1ccc2ccccc2n1`

---

### Kekulé / aromatic

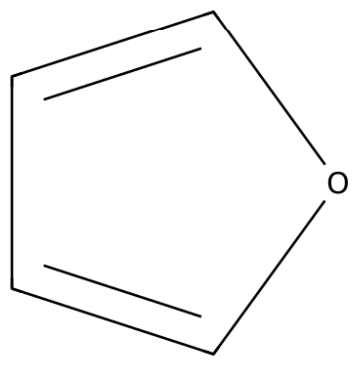

Dataset-provided

`O1C=CC=C1`

RDKit-harmonized

`c1ccoc1`

Supplementary Figure S2. Reconstruction under exact-sequence, connectivity-identity and stereo-aware identity criteria. Full stereochemical identity refers to the annotations encoded in the input, not complete specification of all potential stereochemical sites.

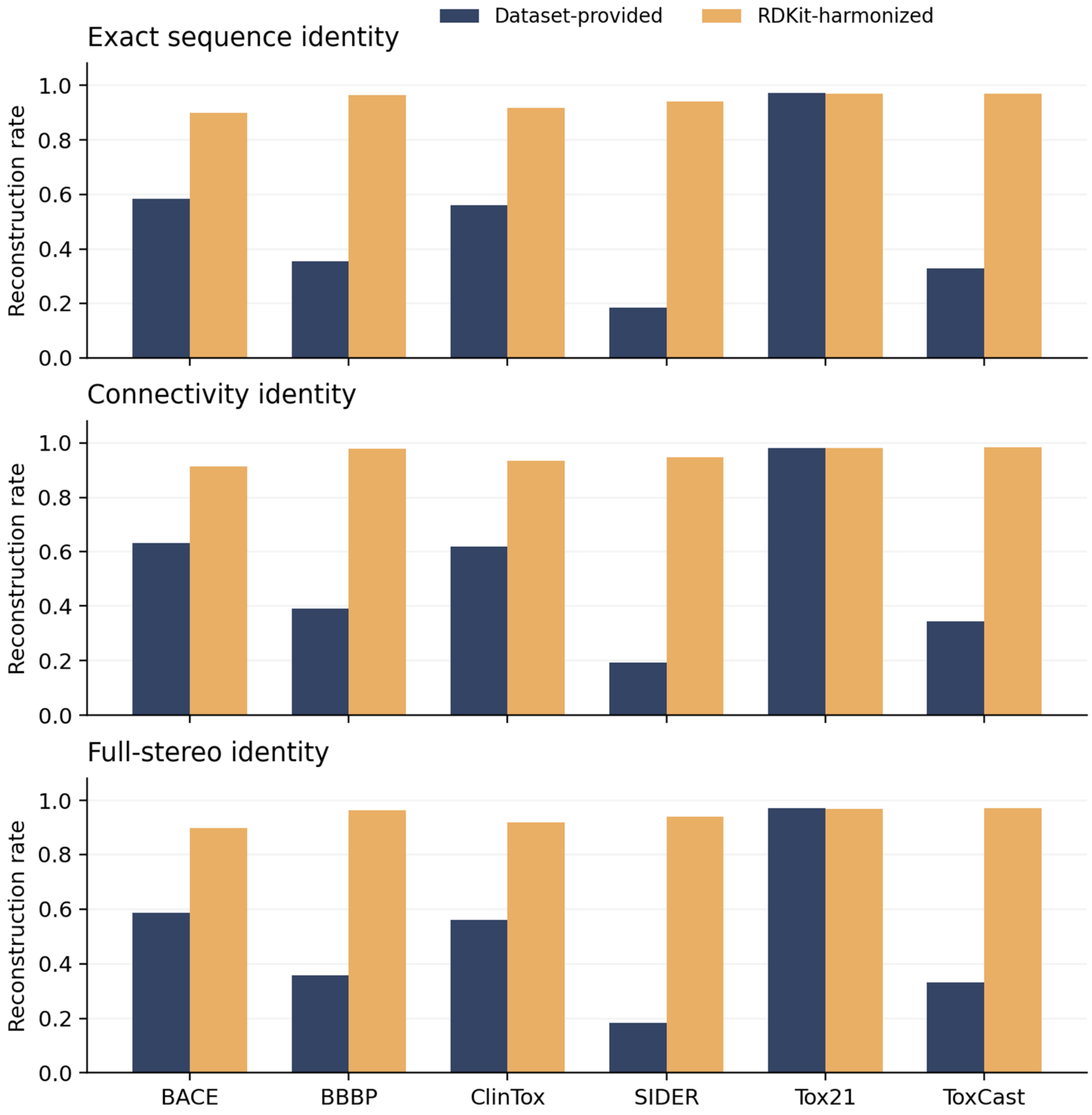

Supplementary Figure S3. Association between normalized Levenshtein distance and paired latent RMS distance among changed same-identity records.

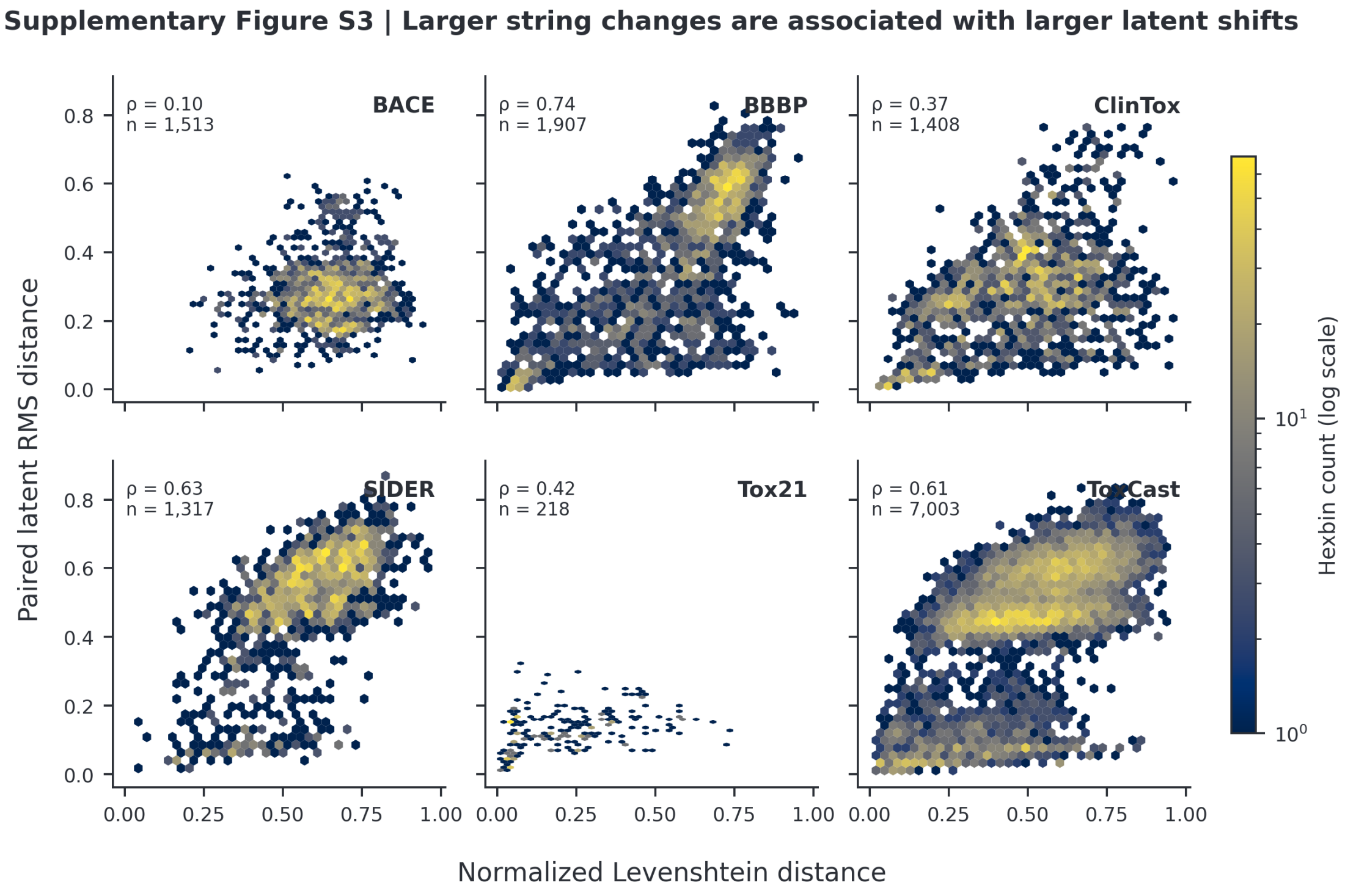

Supplementary Figure S4. Quality-controlled GRU–VAE reconstruction replication. Connectivity-identity reconstruction rates are shown for the three datasets that passed the unchanged-input consistency check. Error bars show 95% Wilson confidence intervals. Δ denotes RDKit-harmonized minus dataset-provided rate; paired difference intervals are reported in Supplementary Table S4.

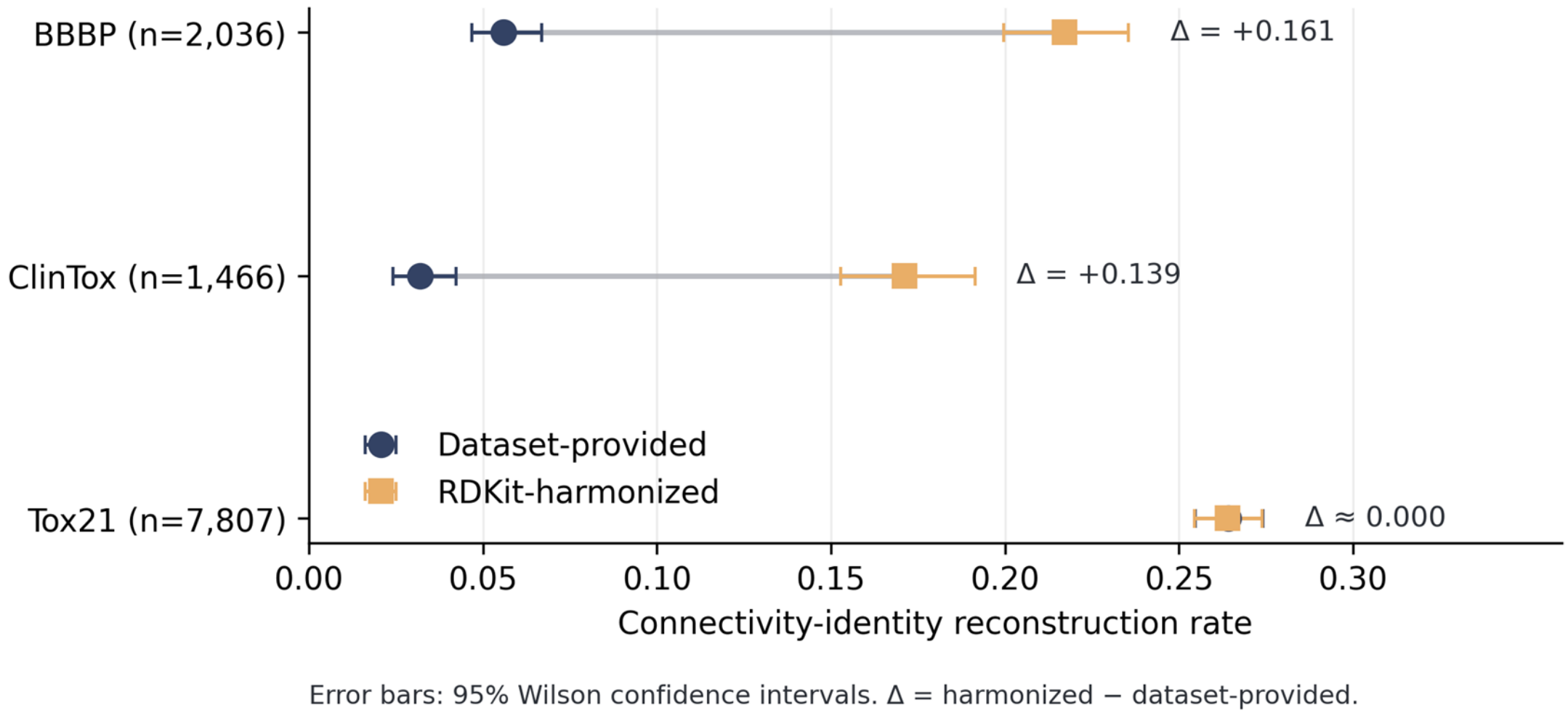

Supplementary Figure S5. Joint t-SNE projections of posterior-mean vectors for BBBP, ClinTox, and SIDER. Dataset-provided and RDKit-harmonized vectors were concatenated and embedded by one t-SNE fit within each dataset. These panels are qualitative.

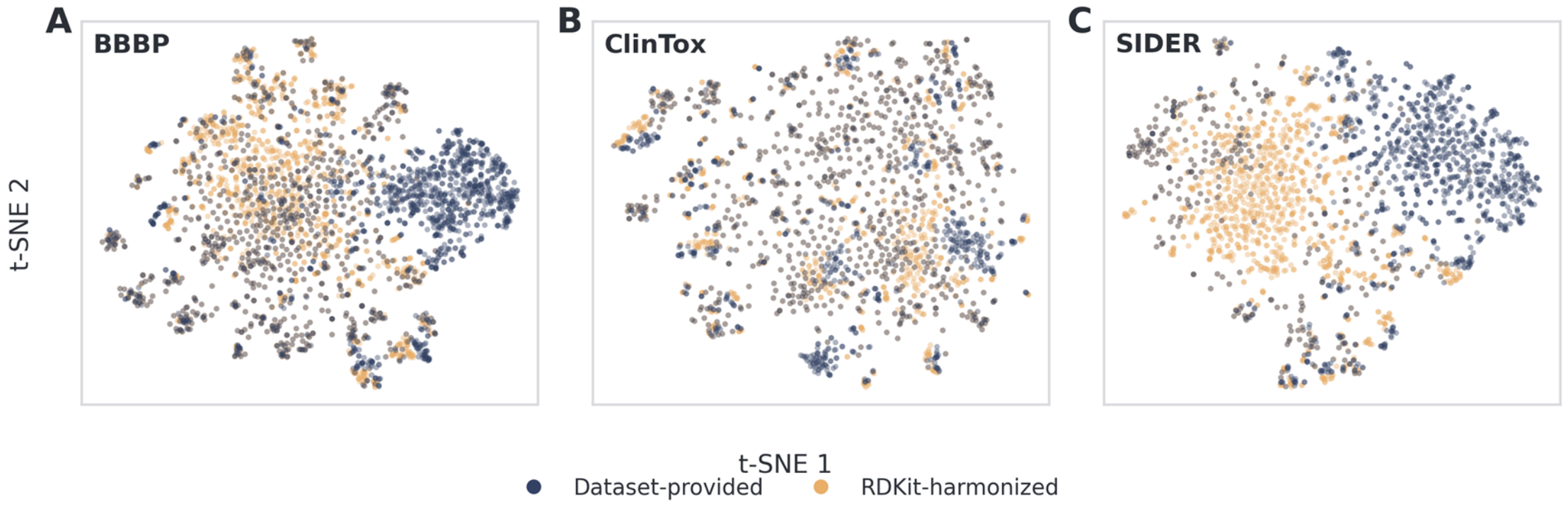

Supplementary Figure S6. Cross-dataset summary of changed-string prevalence, connectivity-identity loss, and median paired latent distance.

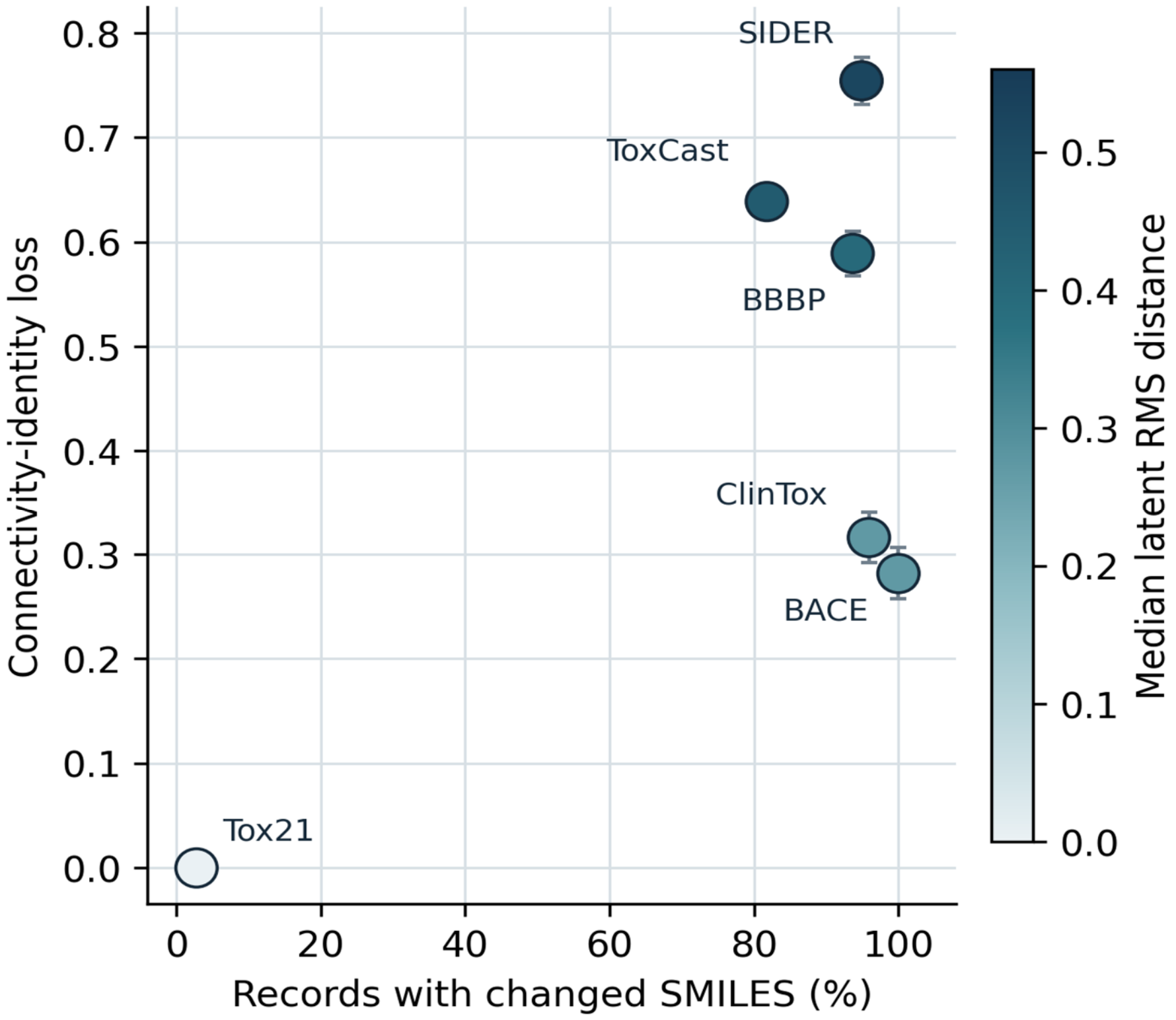

Supplementary Figure S7. Descriptive relation between median paired latent distance and absolute representation-specific performance difference across six datasets.

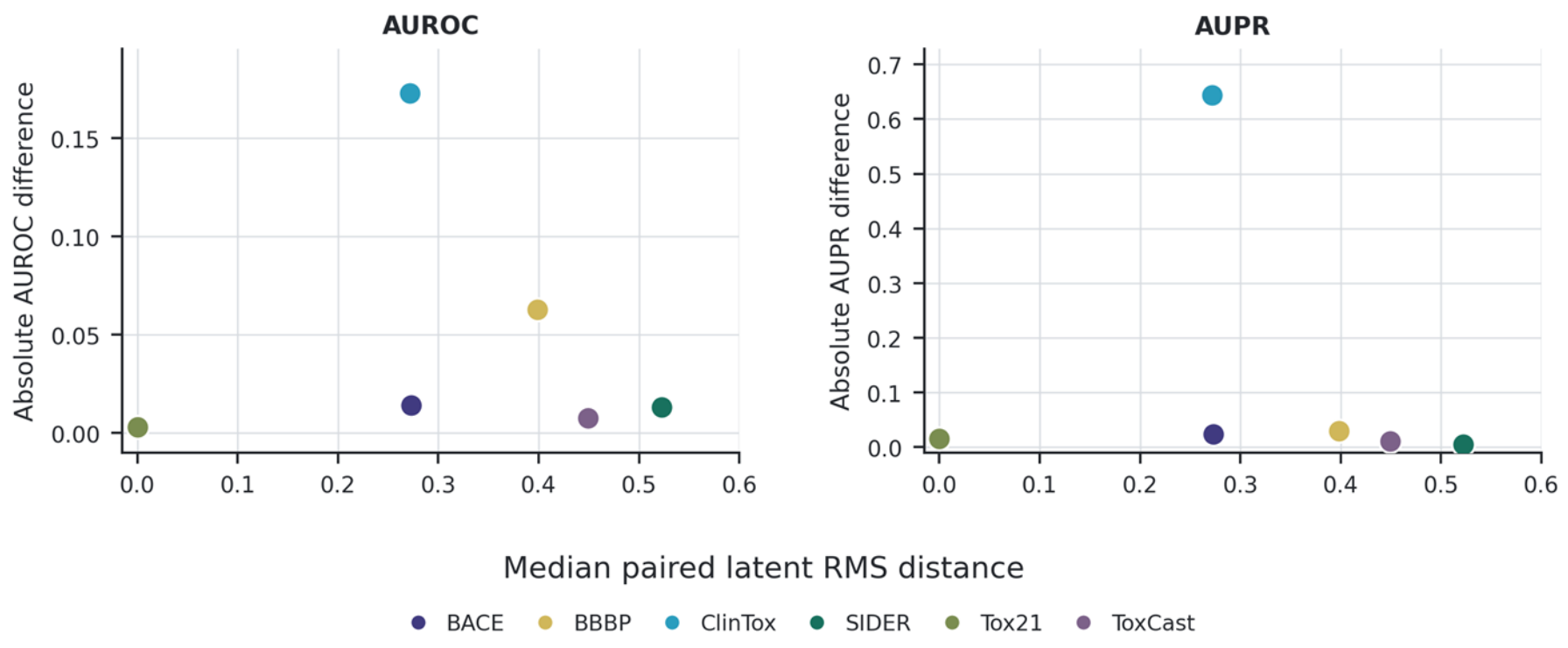

Supplementary Figure S8. K/A enrichment in held-out score rankings. The top row shows ClinTox (n=294; K/A n=17) and the bottom row shows BBBP (n=408; K/A n=178). Left panels show ROC curves, with circles marking true-positive K/A records at their score thresholds. Right panels show cumulative K/A excess and one-sided empirical P values from 10,000 permutations of K/A indicators across fixed score rankings, with a +1 correction. Ties were resolved by test-set row order for ranking and evaluated at their common threshold in the ROC curves.

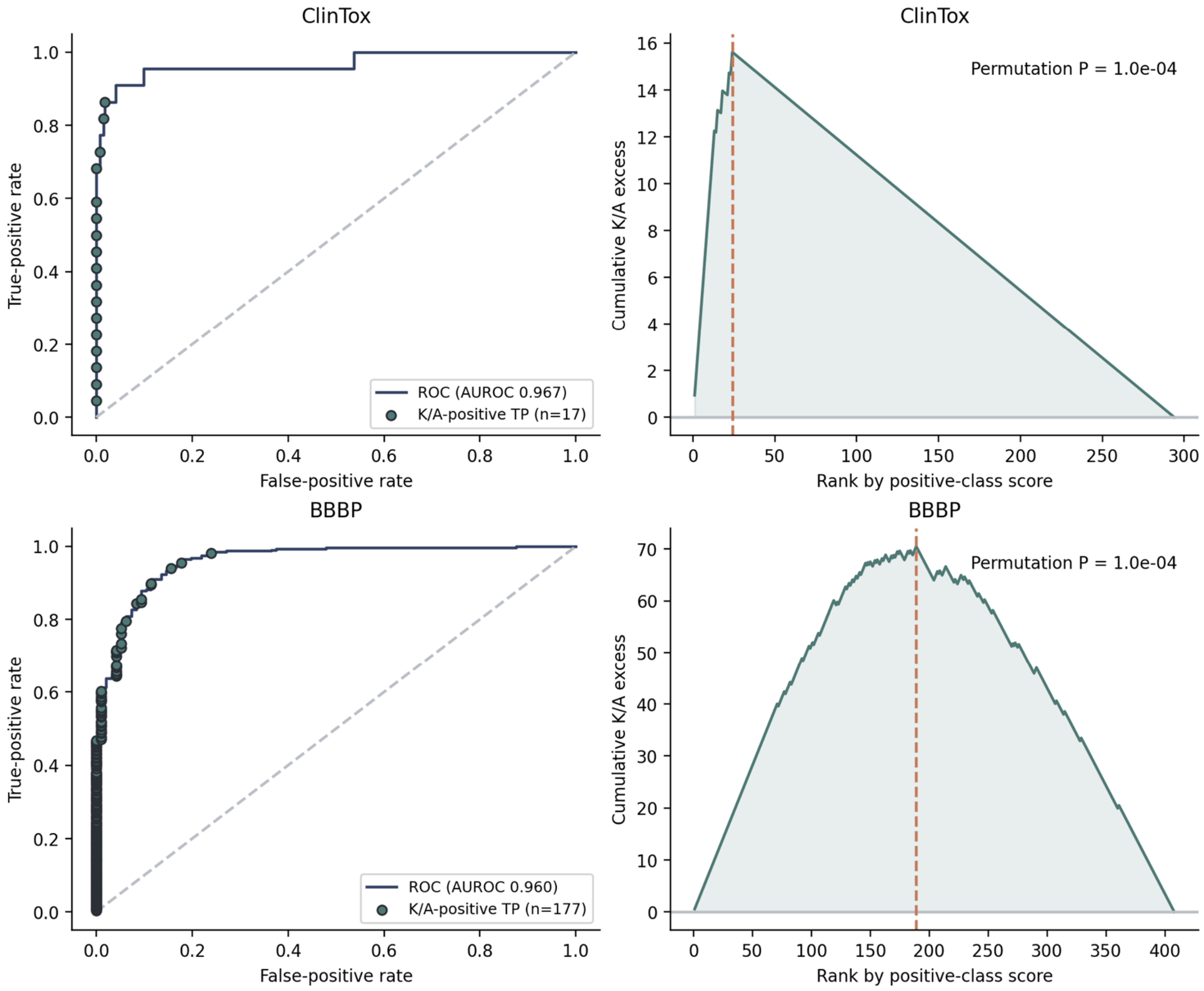